\documentclass{emulateapj}

\newcommand{\Msunpyr}{\ifmmode {\rm\,M_\odot\,yr^{-1}} \else {${\rm\,M_\odot\,yr^{-1}}$}\fi}
\newcommand{\OmM}{\ifmmode {\Omega_{\rm M}}\else $\Omega_{\rm M}$\fi}
\newcommand{\OmL}{\ifmmode {\Omega_{\Lambda}}\else $\Omega_{\Lambda}$\fi}
\newcommand{\kmps}{\ifmmode {\rm\,km~s^{-1}} \else ${\rm\,km\,s^{-1}}$\fi}
\def\HI{{\ion{H}{1}}}
\def\HII{{\ion{H}{2}}}
\def\mum{$\mu$m}

\def\Msun{${\rm M}_{\odot}$}

\def\Ha{H$\alpha$}
\def\Hb{H$\beta$}
\def\OIII{[\ion{O}{3}]}
\def\OII{[\ion{O}{2}]}
\def\OI{[\ion{O}{1}]}
\def\NII{[\ion{N}{2}]}
\def\SII{[\ion{S}{2}]}

\def\OIIILs{[\ion{O}{3}]~$\lambda \lambda$4959,5007\AA}
\def\OIIIL{[\ion{O}{3}]~$\lambda $5007\AA}
\def\OIILs{[\ion{O}{2}]~$\lambda \lambda$3727,29\AA}
\def\OIL{[\ion{O}{1}]~$\lambda $6300\AA}
\def\NIIL{[\ion{N}{2}]~$\lambda $6584\AA}
\def\NIILs{[\ion{N}{2}]~$\lambda \lambda$6548,84\AA}
\def\SIILs{[\ion{S}{2}]~$\lambda \lambda$6717,31\AA}

\def\ratioR23{([\ion{O}{2}]~$\lambda$3727,29+[\ion{O}{3}]~$\lambda\lambda$4959,5007)/H$\beta$}
\def\R23{${\rm R}_{23}$}
\def\S23{${\rm S}_{23}$}
\def\ratioS23{([\ion{S}{2}]~$\lambda \lambda$6717,31+[\ion{S}{3}]~$\lambda\lambda$9069,9532)/H$\beta$}

\received{}
\begin{document}

\title{Modelling the Pan-Spectral Energy Distribution of Starburst Galaxies: III.\\
Emission Line Diagnostics of Ensembles of Evolving \HII\ Regions}

\author{Michael A. Dopita,  J\"org Fischera, \& Ralph S. Sutherland}
\affil{Research School of Astronomy \& Astrophysics,
The Australian National University, \\
Cotter Road, Weston Creek, ACT 2611, Australia}
\author{Lisa J. Kewley,}
\affil{University of Hawaii at Manoa, Institute for Astronomy, 2680 Woodlawn Drive, Honolulu, HI 96822, USA}
\author{Claus Leitherer}
\affil{Space Telescope  Science Institute, 3700 San Martin Drive, Baltimore  MD21218, USA}
\author{Richard J. Tuffs, Cristina C. Popescu, }
\affil{Max-Planck-Institut f\"ur Kernphysik, Saupfercheckweg 1, D-69117 Heidelberg, Germany }
\author{Wil van Breugel}
\affil{Institute of Geophysics and Planetary Physics, Lawrence Livermore National Laboratory, L-413
Livermore, CA 94550, USA}
\author{\& Brent A. Groves}
\affil{Max-Planck Institut  f\"ur Astrophysik, Karl-Schwarzschild-Str 1, 85741, Garching, Germany}
\email{Michael.Dopita@anu.edu.au}


\begin{abstract}
We build, as far as theory will permit, self consistent model  \HII\ regions around central clusters of aging stars. These produce strong emission line diagnostics applicable to either individual \HII\ regions in galaxies, or to the integrated emission line spectra of disk or starburst galaxies. The models assume that the expansion and internal pressure of individual \HII\ regions is driven by the net input of mechanical energy from the central cluster, be it through winds or supernova events. This eliminates the ionization parameter as a free variable, replacing it with a parameter which depends on the ratio of the cluster mass to the pressure in the surrounding interstellar medium. These models explain why \HII\ regions with low abundances have high excitation, and demonstrate that at least part of the warm ionized medium is the result of overlapping faint, old, large, and low pressure \HII\ regions. We present line ratios (at both optical and IR wavelengths) which provide reliable abundance diagnostics for both single \HII\ regions or for integrated galaxy spectra, and we find a number that can be used to estimate the mean age of the cluster stars exciting individual \HII\ regions.
\end{abstract}


\section{\label{intro}Introduction}
Much of what we have learnt about the chemical evolution of the Universe, or of individual galaxies throughout cosmic time has been gleaned from the study of the spatially unresolved emission line specta of distant galaxies \citep{Steidel96,Kobulnicky99a,Kobulnicky99b,Kewley05}. This emission line spectrum arises from the ensemble average of \HII\ regions within the galaxy, characterized by a wide range of physical parameters.

For example, in disk galaxies, the chemical abundance of heavy elements falls continuously from centre to edge. In like manner, both the pressure and density of the interstellar medium in the disk drops exponentially with radius. For a given input of mechanical energy by the central star cluster, this means that the outermost \HII\ regions expand more rapidly, and are both larger and have lower internal pressure at a given age. However, the properties of the central cluster of a given mass change systematically with chemical abundance. Massive stars with higher chemical abundances lose more mass prior to supernova explosion, have stronger stellar winds, evolve more rapidly, and have lower effective temperatures while they are on the Main Sequence. 

Finally, the temporal evolution of individual \HII\ regions is extremely important in determining the integrated emission line spectrum. \citet{Leitherer92} demonstrated the importance of aging of the OB stars in lowering the number of ionizing photons as the most massive stars evolve away from the Zero Age Main Sequence (ZAMS) to become supergiants. Later, these stars enjoy a brief resurgence in their ionizing photon and mechanical energy production as they become Wolf-Rayet stars.

With so many temporal and physical variables, it is perhaps not surprising that the nebular modelling community (ourselves included) have in the past been reduced to making severe, even gross, oversimplifications in attempts to model the integrated strong emission line spectra of galaxies. Indeed, frequently only a single line ratio has been used to determine the metallicity of extragalactic \HII\ regions or even of whole galaxies. This is the famous  \R23 ratio; \ratioR23.

This ratio was first proposed by \citet{Pagel79}. The logic for its use is impeccable, since it uses the two strongest lines of the strongest coolant of \HII\ regions, and it should therefore be sensitive to the total oxygen abundance. By using two stages of ionization, it accounts for the emission in the bulk of the ionized nebular volume. However, the calibration of this ratio in terms of the abundance has proved to be very difficult and many different and often contradictory calibrations of \R23 are available, including \citet{Pagel79,Pagel80,Edmunds84,McCall85,Dopita86,Torres-Peimbert89,Skillman89,McGaugh91,Zaritsky94,Pilyugin00,Kewley02}.

The basic cause of this difficulty in calibrating the \R23 ratio is common to many ratios formed from the flux of optical forbidden line divided by the flux of a recombination line of hydrogen. Initially, as $Z$ is increased, the forbidden line increases in flux, and the line ratio increases. However, as the abundance increases further, the cooling by forbidden lines lowers the electron temperature making it more difficult to collisionally excite the optical forbidden lines. Eventually there comes a point at which an increase in abundance is matched by a decrease in collisional excitation and the line ratio reaches its maximum. Any further increase in $Z$ then leads to a \emph{decrease} in the relative strength of the forbidden line. \R23, like other optical diagnostics is therefore a two-valued function of $Z$. Because infrared lines have lower thresholds for collisional excitation, they remain monatonic functions of $Z$ up to much higher metallicities. Recognizing this physics, the S$_{23}$ ratio; \ratioS23, or even more complex ratios, have been used either in the place of or to supplement the \R23 ratio \citep{Dennefeld83,Kennicutt96,Oey02}.

The calibration problems of the \R23 ratio are made worse by the scatter, at fixed  \R23 and within each given branch, both between models and observational data, and amongst different models. The observational material is affected by uncertainty in the reddening correction applying to the \OIILs\ , since classical extinction laws are usually used, rather than the attenuation laws which more correctly apply to extended objects \citep{Fischera03,Fischera05}. This point will be discussed more fully below. As far as the theoretical models are concerned, scatter in the relationship is easily generated, since the stellar atmospheres used for single stars can show very wide differences in the number of ionizing photons produced per unit frequency \citep{Morisset04}.

Moving beyond the use of single line ratios, a more sophisticated analysis is made possible by the understanding that the strong line emission spectrum of an individual \HII\ region is controlled by three physical parameters \citep{Dopita86}. These parameters are, the effective temperature of the exciting stars, the chemical abundance set (or ``metallicity", $Z$) and the ionization parameter, defined either as the ratio of the mean ionizing photon flux to the mean atom density, $q$ or in its dimensionless form as the ratio of mean photon density to mean atom density ${\cal{U}}=q/c$. Thus, for a given stellar input spectrum a ${\cal{U}} : Z$ grid of models should provide the ideal means of interpreting the observational data \citep{Kewley02}.

This analysis leaves open the issue of the age distribution of the exciting stars. The metallicity and the age of the exciting stars determines the energy distribution of the EUV photons. In the modelling hitherto, this problem is usually treated in two limits;  the instantaneous burst approximation (for which zero age is often, but not universally assumed) or the continuous star formation approximation in which stars are assumed to be born continually within the ionized region (see \citet{Kewley01a}, and references therein). All this modeling  effectively assumes a single \HII\ region of a given metallicity and ionization parameter. Whilst such an approach is acceptable in attempts to understand the emission line spectrum of individual  \HII\ regions, it is certainly inadequate in attempts to interpret the integrated spectrum of a whole galaxy, or even of a complex of many \HII\ regions. 

The advent of the  \emph{Starburst 99 v.5} \citep{Leitherer99} stellar spectral synthesis code finally allowed the construction of fully self-consistent models for \HII\ regions. For a given stellar initial mass function (IMF), this code delivers not only the pan-spectral distribution of photons, including ionizing photons as a function of age, but also provides tables of the time-dependent mechanical energy input of the stars including OB-star stellar winds, red giant winds, Wolf-Rayet winds and the energy input from supernova explosions.

In the first paper of this series (\citet{Dopita05}, hereinafter referred to as Paper I), we computed the (one dimensional) evolution of \HII\ regions around clusters of a single mass. We demonstrated how the size~:~age relationship is essentially determined by the density (or pressure) in the surrounding interstellar medium (ISM). This in turn determines the dust `temperature' described by the wavelength of the peak of the far-IR dust re-emission feature.

In the second paper of this series, Dopita et al. (2006), in press, hereinafter referred to as Paper II), we investigated the role of metallicity and of the cluster mass function in controlling the excitation of the populations of \HII\ regions in galaxies, be they disk or starburst. 

We demonstrated that the ionization parameter of \HII\ regions, previously treated as a free variable, is determined at any time by the instantaneous ratio of the ionizing photon flux to the mechanical energy flux of the central stars. This is because the HII region is evolving as a mass-loss bubble pressurized by the combined ram pressure of the stellar wind and supernova explosions, so HII region density and HII region radius are closely coupled. Although this conclusion was obtained using a simple 1-D spherical evolution model for the \HII\ region, it should remain valid in more complex geometries, since wherever un-ionized or molecular inclusions exist, these will be pressure confined by the hot shocked stellar wind gas, and this gas pressure is coupled with the overall size of the \HII\ region, which itself determines the dilution of the radiation field of the central stars.

Because the effective temperature of the exciting stars increases as metallicity is lowered, but the mechanical energy flux in the stellar winds decreases towards to lower metallicity, low abundance HII regions are characterized by higher ionization parameters, a fact that had been frequently noted by observers, but which had not previously been explained in a satisfactory manner. 

Using these insights, in this third paper of the series,  we proceed to the next stage of sophistication in modeling -- the construction of time-dependent photoionization models of individual \HII\ regions, and the construction of ensemble averages of aging \HII\ regions to provide galaxy-wide averaged spectra at a given metallicity.  Such models applied to individual \HII\ regions make it possible to estimate the ages  of the exciting stars from the positions of the observations on theoretical \HII\ region isochrones. Previously, this could not be done, since the ionization parameter and the effective temperature of the cluster  (determined by the ageing of its stars) can both produce similar effects on the emitted spectrum. This degeneracy is raised thanks to the self-consistent geometry that we have developed, which includes a dynamical evolution of the \HII\ region consistent with the properties of the stellar wind generated by the central cluster.

In this paper, we apply these new techniques to produce line ratio diagnostics which enable the independent determination of both the stellar ages and metallicities of the exciting clusters for individual \HII\ regions. These should prove useful in the analysis of abundance gradients in resolved disk galaxies. More importantly, we develop line ratio diagnostics for the determination of abundances in ensembles of \HII\ regions in starburst and normal galaxies. This should greatly assist in analyses of the strong emission lines of the more distant and unresolved galaxies, observed either at optical wavelengths or in the IR. 

\section{\label{models}Models}
\subsection{Codes}
We have used the \emph{Starburst 99 v.5} code in its latest (2005) version to compute the pan-spectral distribution of clusters of stars with a piece-wise fit to a Miller-Scalo initial mass function (IMF) \citet{M-S79} between 0.1 and 120 \Msun. The parameters of this fit are the same as given in \citet{M-S79} and were also tabulated by us in Paper II. Either the use of a Miller-Scalo IMF or of a Salpeter IMF would not noticeably affect the line ratio diagnostics presented in this paper, since the slopes of the IMF are almost identical above 10 \Msun. However, the choice of the IMF has a much larger effect on the total mass of the cluster, since this is determined by the choice of the power law below 10 \Msun.

Stellar atmosphere models for stars with plane parallel atmospheres are based on the \citet{Kurucz92} models as  compiled by \citet{Lejeune97}.  The fully line-blanketed Wolf-Rayet atmosphere models of \citet{Hillier98}  and the non-LTE O-star atmospheres of \citet{Pauldrach01} have been incorporated into \emph{Starburst 99 v.5} as described in \citet{Smith02}. The flux distributions following from these
atmospheres are coupled to the stellar evolution models by the Geneva group. We used the so-called "high-mass-loss" tracks, which provide the best match to the observed stellar inventory in the Hertzsprung-Russell diagram \citep{Leitherer99}.

For the Initial Mass Function (IMF) we used piece-wise power-law fits to the \citet{M-S79} mass function, although we note that their paper, \citet{M-S79}  represented their IMF as a truncated lognormal distribution (with its maximum at zero). However, they also represented it as a three-segment broken power law, which is the form of the IMF which \emph{Starburst 99} currently accepts. The logarithmic slopes of the piecewise power-law fits, $\Gamma$, and the mass range applicable to each segment were given in Table 1 of Paper II.

For a given cluster mass, we used the stellar wind and supernova power output, as tabulated by the \emph{Starburst 99 v.5} code, to solve for the radius of the \HII\ regions as a function of time given the pressure or, equivalently, the mean density in the ISM. This requires a Runge-Kutta integration of the equation of  motion of the swept-up shell, as was done and described in both Paper I and Paper II.

The pressure in the \HII\ region is the same as in the shocked stellar wind gas. This is given by the classical \citet{Castor75} theory;
\begin{equation}
P(t) = {\frac {7 } {(3850\pi)^{2/5}}}\left(\frac{250}{ 308\pi }\right)^{4/15}{\left(\frac {L_{\rm mech}(t)}{ \mu m_Hn}\right)}^{2/3} {\frac{\mu m_H n_0}{r(t)^{4/3}}} \label{P1} 
\end{equation}
where $L_{\rm mech}(t)$ is the instantaneous production of mechanical energy by the cluster stars, $r(t)$ is the radius, $n_0$ is the mean atom density in the surrounding ISM and $n$ is the mean atom density in the ionized gas and $\delta(t)=n/n_0$ is the instantaneous ratio of the density interior to the bubble to the density in the surrounding ISM.

In Paper II, we showed that the ionization parameter at the contact discontinuity in the swept-up shell of ISM, which can be treated as the inner boundary of the \HII\ region depends largely on the instantaneous properties of the exciting cluster stars;
\begin{equation}
q(t) \propto \delta(t)^{3/2}S_*(t)/L_{\rm mech}(t). \label{q}
\end{equation}
where $S_*(t)$ is the instantaneous flux of ionizing photons from the central cluster.

The presence of the $\delta(t)$ factor in the above equation provides a weak coupling between the ionization parameter and both the pressure in the ISM, $P_0$, and the mass of the central cluster, $M_{cl}$. Together, these determine the strength of the outer shock of the mass-loss bubble and therefore the compression factor through it. The appropriate scaling factors are  $q \propto P_0^{-1/5}$ and  $q \propto M_{cl}^{1/5}$. Thus, once $P_0$ and $q \propto M_{cl}^{1/5}$ are fixed, the ionization parameter is determined.

Using this dynamical model to derive the instantaneous size and ionization parameter of the \HII\ region, the latest version of our code \emph{Mappings IIIr} \citep{Sutherland93,Dopita02,Groves04} with the physical parameters defined in Paper I was used to compute photoionization models. The resulting model isobaric dusty \HII\ regions were used to produce a table of emission lines, described in the Appendix, and given fully the Electronic version of the paper. These enable us to compute the line ratio diagnostics described in this paper.

\subsection{Metallicities \& Depletion Factors}

The solar abundance set adopted here is taken from \citet{Asplund05}. This incorporates the results of many papers which together have provided a self-consistent re-calibration of the solar abundance. In many cases, these abundances are nearly a factor of two lower than those used previously \citep{Anders89}. The corresponding logarithmic abundance ratios of the elements with respect to Hydrogen are given in table \ref{table1}. For the purposes of the modeling, when we refer to solar metallicity, $Z_{\odot}$, we are using this set of abundances. 

We note here that as a consequence of the re-scaling of the solar abundances (which is not yet reflected in the stellar models), the \emph{Starburst 99} ``solar" abundance set is inconsistent with the new \citet{Asplund05} abundance set. The  \emph{Starburst 99} code uses absolute metallicities of $Z=0.001, 0.004, 0.008, 0.02$ and 0.04, which makes the   \emph{Starburst 99} ``solar" metallicity, $Z=0.02$, higher than the \citet{Asplund05} ``solar metallicity" , $Z=0.016$. Whilst this  matter very little in terms of the number of ionizing photons emitted by the stars, it will mean that the computed stellar UV photon field will slightly softer than it should be, given the nebular abundance set.

The depletion factors from the gaseous phase determine the composition of the dust in the models. For these we have took the measured gas-phase abundances in the local interstellar cloud by \citet{Kimura03}, and used these to infer a set of local depletion factors consistent with the  \citet{Asplund05} abundance set. The adopted depletion set is also given in table \ref{table1}. For models with non-solar abundances, the depletion factors were held constant at solar values. This is equivalent to taking the dust~:~gas ratio to be proportional to metallicity.

Three elements do not scale simply with the metallicity. First, because of its high initial abundance as a result of nuclear burning in the Big Bang, Helium scales only weakly with metallicity. For this element,
we assume a primary nucleosynthesis component in addition to its primordial value. From \citet{Russell92} and \citet{Pagel92} we infer the empirical relationship:
\begin{equation}
\rm{He/H} = 0.0737+0.024\times {Z/Z_{\odot}}.
\end{equation}

For both Nitrogen and Carbon there is clear evidence that these are both a primary nucleosynthetic element, dominant at low metallicity, and a secondary nucleosynthetic element once higher abundances are reached. Alternatively, both are produced in part by dredge-up in intermediate mass stars, causing a delay in the onset of enrichment of these elements from this source. For Nitrogen we have used a modified version of the form proposed by \citet{Groves04};
\begin{equation}
{\rm{N/H}}=1.1\times 10^{-5} {Z/Z_{\odot}} +4.9\times 10^{-5} {Z/Z_{\odot}}^2.
\end{equation}

\begin{figure}
 \includegraphics[width=\hsize]{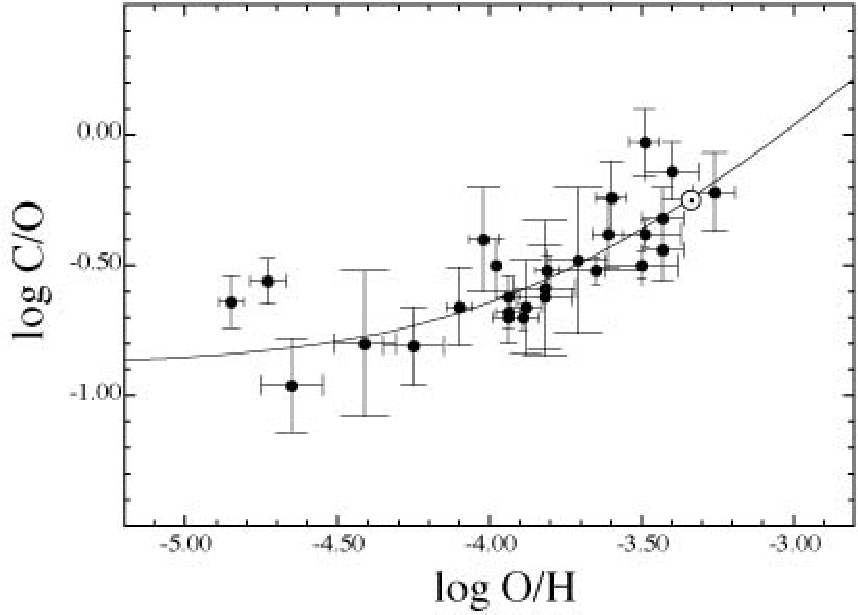}
  \caption{\label{fig1}
The fit given by equation \ref{eq3} is shown along with the HST observations of \HII\ regions made by \citet{Garnett99}. The solar abundance from \citet{Asplund05} is also marked with a $\odot$ symbol. The new solar abundance set eliminates the systematic difference between the solar and the \HII\ region abundances which was noted by Garnett et al.}
\end{figure}

For Carbon, we have used the HST data on the Carbon abundances of \HII\ regions as compiled by \cite{Garnett99}. This provides a fit of the same mathematical form as for N:
\begin{equation}
{\rm{C/H}}=6.0\times 10^{-5} {Z/Z_{\odot}} +2.0\times 10^{-4} {Z/Z_{\odot}}^2. \label{eq3}
\end{equation}
The fit to the observations is shown in figure \ref{fig1}.

\begin{deluxetable}{lrr}
\tabletypesize{\small}
\tablewidth{0pt}
\tablecaption{Adopted solar chemical abundance ratios, $\log[n_{\rm X}/n_{\rm H}]$, and depletion factors (D).\label{table1}}
\tablehead{
\colhead{Element}
& \colhead{$\log \left[ N_{\rm X}/n_{\rm H} \right]$}
& \colhead{$\log({\rm D})$}\\
}
\startdata
H & 0.00 & 0.00 \\
He & -1.01 & 0.00 \\
C & -3.59 & -0.15 \\
N & -4.22 & -0.23 \\
O & -3.34 & -0.21 \\
Ne & -3.91 & 0.00 \\
Mg & -4.47 & -1.08 \\
Si & -4.49 & -0.81 \\
S & -4.79 & -0.08 \\
Ar & -5.20 & 0.00 \\
Ca & -5.64 & -2.52 \\
Fe & -4.55 & -1.31 \\
\enddata
\end{deluxetable}

\subsection{Parameters of the Photoionization Models}
The \emph{Starburst 99 v.5} code allows us to study line ratios at five (fixed) absolute metallicities of $Z=0.001, 0.004, 0.008, 0.02$ and 0.04. For each of these abundance sets we have computed several families of \HII\ region models with ages 0.2, 0.5, and 1.0~Myr and in further 0.5~Myr steps up to a maximum age of 6.5~Myr. By this time, more than 96\% of ionizing photons have been emitted, the \HII\ regions have faded by factors of between 5 and 12 from their peak luminosity, and their surface brightnesses will have faded by much greater factors than these.

The ionization parameter is not a free variable in these models. The geometry of the gas with respect to the ionizing stars is instead determined by the equation of motion of the expanding \HII\ region under the driving pressure of the stellar mass-loss and supernova explosions. This pressure also determines the density of the ionized plasma. As was shown in Paper II, the equation of motion and the pressure together lead us to infer that there is a weak coupling between the ionization parameter and both the pressure in the ISM, $P_0$, and he mass of the central cluster, $M_{cl}$. Together, these determine the strength of the outer shock of the mass-loss bubble and therefore the compression factor through it. The appropriate scaling factors are  $q \propto P_0^{-1/5}$ and  $q \propto M_{cl}^{1/5}$. 

\begin{figure}
\includegraphics[width=\hsize]{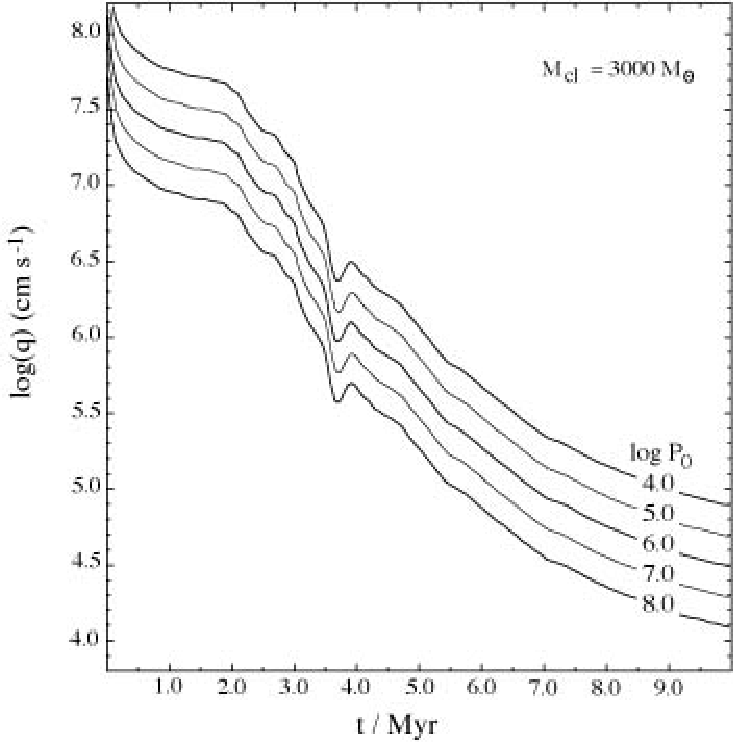}
  \caption{\label{fig2}
The computed run of ionization parameter at the inner edge of the \HII\ region as a function of time and ISM pressure $log P_0/k $ (cm$^{-3}$K) for a cluster having a mass of $3 \times 10^3M_{\odot}$ and assuming a Miller-Scalo IMF. A given choice of  the variable ${\cal R} = (M_{\rm cl}/M_{\odot})/(P_0/k)$ defines a unique run of the ionization parameter with time.}
\end{figure}

The ratio, ${\cal R} = (M_{\rm cl}/M_{\odot})/(P_0/k)$, with $P_0/k$, measured in cgs units (cm$^{-3}$K), uniquely determines the run of ionization parameter with time. It is this variable that replaces the ionization parameter and other geometrical considerations in our models. In figure \ref{fig2} we show the run of the computed ionization parameter $q$ as a function of time and ISM pressure for a solar abundance cluster with $M_{cl}=3 \times 10^3M_{\odot}$ and having a Miller-Scalo IMF. The (dimensional) ionization parameter $q$ is related to the more frequently used (dimensionless) ionization parameter $\cal{U}$ by ${ \cal U} = q/c$. 

The run of $q$ with time is computed using the \emph{Starburst 99} output, and solving for the time dependent radius and internal pressure of the mass-loss bubble as described above. In the early phases ($t < 2$~Myr) the ratio of ionizing photon flux to mechanical energy flux is almost constant, and the changes in $q$ mostly reflect the size evolution of the mass-loss bubble. After 2~Myr, the ionizing photon flux rapidly decreases as massive stars evolve to red supergiants, and then the mechanical energy flux increases as a result of Wolf-Rayet stellar winds. Both of these lead to a strong decline in $q$. At around 3.5~Myr, supernova explosions add to the internal pressure of the bubble, sharply decreasing the $q$ in the ionized gas and re-accelerating the expansion of the mass-loss bubble. At later times, the progressive death of the high-mass stars leads to further declines in the ionizing photon flux, but the decline in $q$ is slower because the internal pressure of the mass-loss bubble also declines, partially offsetting the decline in ionizing photon flux.

The theoretical ionization parameter rather strongly dependent upon the chemical abundance. It is driven by two factors. First, at higher abundance, the stellar wind has a higher opacity and therefore absorbs a greater fraction of the ionizing photons, reducing the $q$ in the surrounding \HII\ region. Second, the atmosphere scatters the photons emitted from the photosphere more efficiently when the atmospheric abundances are higher, leading to a greater conversion efficiency from luminous energy flux to mechanical energy flux in the stellar wind base region. This also leads to a diminution of  $q$ in the surrounding \HII\ region. These factors acting together provide a sensitivity to metallicity of  $q \propto Z^{-0.8}$, approximately.

As described above at any age $t$, the instantaneous ionization parameter - which determines the excitation of the \HII\ region -  is governed by the ratio of the mass of the cluster and the pressure in the surrounding ISM. Since cluster masses may vary between $\sim 100 M_{\odot}$ and $10^6 M_{\odot}$, and the likely range of ISM pressures are between $10^4 < P_0/k  < 10^7$ cm$^{-3}$~K it follows that reasonable values for $\cal{R}$ fall within the range $-6 < \log {\cal R} < +2$. We have therefore computed, for each metallicity, an age sequence of model \HII\ regions having $\log {\cal R} = -6, -4, -2, 0$ and +2. For a given metallicity, the instantaneous value of the ionization parameter depends on ${\cal{R}}^{1/5}$, as we have shown in Paper II. Note that a particular choice of ${\cal R}$ is not tied to any particular cluster mass. Thus, a cluster of $10^5 $ M$_\odot$ would be characterized by $ \log {\cal R} \sim 0.6$ in the solar vicinity, and $ \log {\cal R} \sim -1.0$ in a high-pressure starburst environment.

It is worth noting at this point that this parameterization of the problem of evolving \HII\ regions only remains valid for as long as stochastic variation in the cluster membership may be ignored. This means that the exciting cluster has to be massive enough that the upper IMF is adequately populated with stars so as to avoid stochastic variations in either the mass to light ratio or the effective temperature of the cluster stars. This effectively limits the cluster mass for which these models are valid,  $M_{\rm cl} \ge10^3 M_{\odot}$.

Together, the two parameters ${Z/Z_{\odot}}$ and $\log {\cal R}$ define a unique \HII\ region spectrum, with one caveat, namely, that if the density of the ionized gas becomes too high, then collisional de-excitations of forbidden lines will alter the emergent spectrum. This only becomes a concern at the very highest values of ISM pressure either within ultra-compact \HII\ regions (as computed by \citet{Dopita06} or else in starburst environments with $P_0/k > 10^7$ cm$^{-3}$~K. For the current purposes, we have computed the particular case of $P_0/k = 10^6$ cm$^{-3}$~K, which constrains the electron density in the \HII\ region below $\sim 100$cm$^{-3}$ except at very earliest times and in the models of highest metallicities, where it may reach  $\sim 10^3$cm$^{-3}$.

For each model, we have prepared tabular data for all the strong lines between the Lyman Limit and 88\mum, from which we form H$\beta$ flux weighted time averaged spectra to represent the integrated spectrum for a given $Z/Z_{\odot}$ and $\log {\cal R}$. The set of strong spectral line intensities with respect to \Hb\ are available on-line. The form of the tables is given in the Appendix, Tables 2 - 7. The code also returns the intensities of very many more fainter lines, which are not listed here.

\section{\HII\ Region emission Line Diagnostics}
\subsection{Optical Lines}
\subsubsection{Veilleux \& Osterbrock Diagnostics}
The diagnostics that have been most frequently used in the classification of the nature of nebular excitation are those proposed by \citet{Veilleux87} (hereinafter V\&O). In these, the \NIIL/H$\alpha$ the \SIILs/ \Ha\ or the [\ion{O}{1}]$\lambda6300$/\Ha\ \emph are plotted against the \OIIIL/H$\beta$ ratio.

These V\&O ratios have the virtue of only using lines which are close together in wavelength in forming the ratio, so errors due to uncertain reddening corrections are avoided. These plots also nicely separate \HII\ regions from regions excited by active galactic nuclei (Seyfert or LINER nuclei) and from shock-excited objects. The \HII\ regions form an extremely  tight sequence, particularly in the diagram using the \NIIL/\Ha\ ratio.

\begin{figure*}
 \includegraphics[width=\hsize]{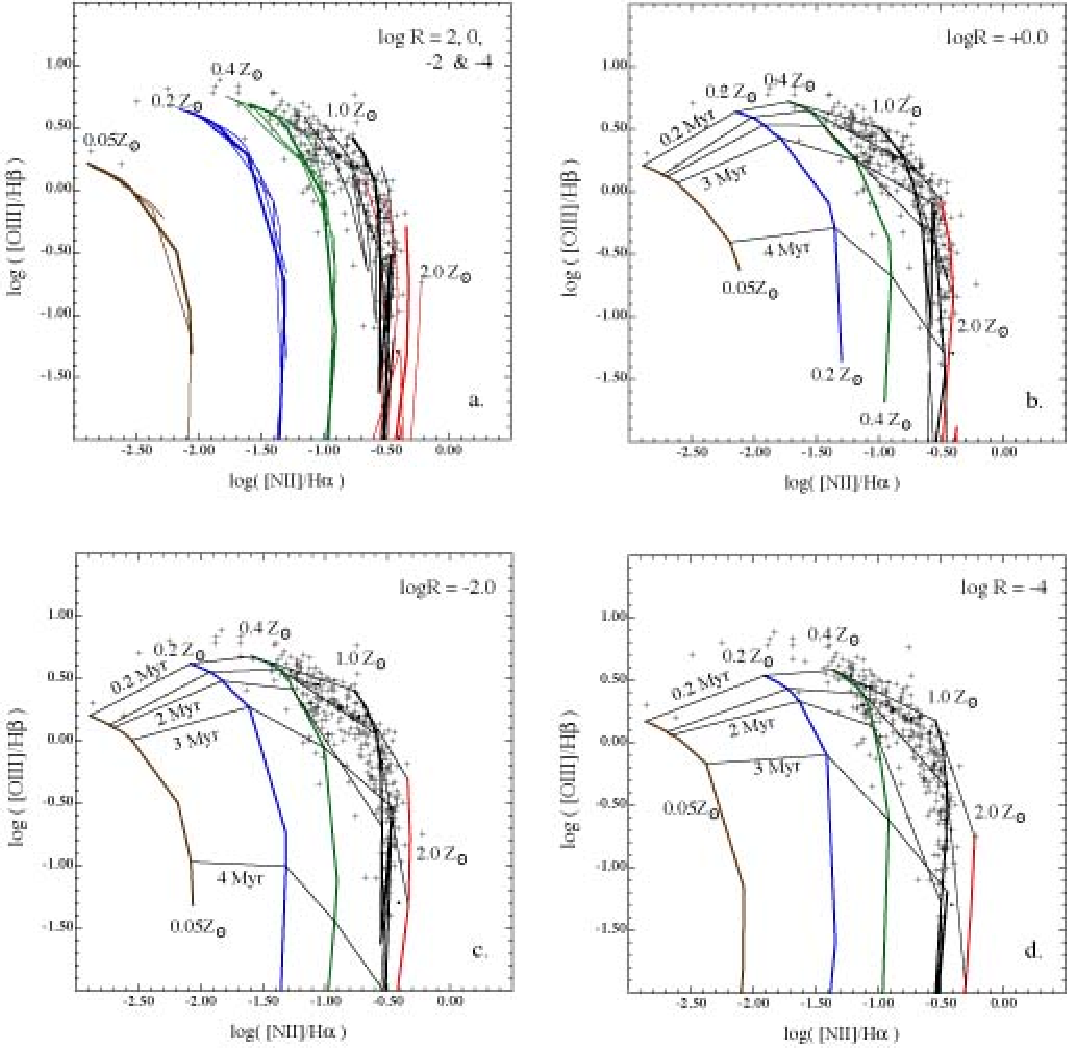}
  \caption{\label{fig3}
The  \NIIL/\Ha\ \emph{vs.} the \OIIIL/\Hb\ ratio. The abundances are coded by color; for $Z/Z_{\odot} 0.05, 0.2, 0.4, 1.0$ and 2.0 the colors are brown, blue, green, black and red, respectively. In panel (a) we plot all models of a given metallicity, and in the remaining panels we plot ageing tracks at each metallicity and the corresponding isochrones for three likely values of the $R$ parameter; $\log {\cal R} = 0, -2$ and -4, respectively.}
\end{figure*}

The results of our modelling for the \NIIL/\Ha\ \emph{vs.} the \OIII/\Hb\ ratio is shown in figure \ref{fig3}. Because of the difficulty of representing so many theoretical models on the same figure, this is presented here in color, and in four panels. The first panel shows models of all ages for each metallicity, color coded for clarity. The model series corresponding to $\log {\cal R}=-2$ is drawn bold, since this is our best guess of the most likely value of $\log {\cal R}$ in disk galaxies ($10^2 \leq M_{\rm cl}/M_{\odot} \leq 10^4$ and $10^4 \leq P/k \leq 10^6$~cm$^{-3}$K). The remaining three panels show the complete grid of models for $\log {\cal R} =0, -2$ and -4, respectively. Isochrones are plotted for 0.2, 1.0, 2.0, 3.0 and 4.0~Myr.

In this figure, the observational datasets for individual \HII\ regions are presented as crosses. The data are drawn from \citet{vanZee98, Kennicutt96,Dennefeld83,Walsh97,Roy97} and the extension of the data set to low abundances is from \citet{Pagel92}.

The reason why the  observed \HII\ regions on a \NII/\Ha\  \emph{vs.} the \OIII/\Hb\ form such a narrow sequence is quite evident. First, to be bright, and therefore selected for observation, observed \HII\ regions should have an age less than $\sim 3$~Myr. In this age range, the isochrones of all models of different abundances fall into the very narrow strip defined by the observations. Although it depends somewhat on the assumed value of $\log {\cal R}$, it is clear that no observed \HII\ region has an age greater than $3-4$~Myr, consistent with the presumption of strong observational selection against old \HII\ regions.

This figure should be compared with Figure 2 of \citet{Dopita00}, in which a grid of models in $Z/Z_{\odot}$ and $q$, the ionization parameter were presented. It is clear that, in order to have a low metallicity \HII\ region with a low ionization parameter, the \HII\ region has to be faint and old, and would therefore not have been observed in these surveys. For metallicities greater than about  $Z/Z_{\odot}\sim 0.5$, the models are highly degenerate in terms of both the metallicity and the cluster age. This accounts for the very tight distribution of observed points in the region of the plot where  $\log$(\OIII/\Hb)$ < 0.0$.

Since the models of \citet{Dopita00} were run, the EUV field predicted by the \emph{Starburst 99 v.5} code has softened, thanks to improvements in the treatment of extended atmospheres. This leads both to the reduction in the limiting \NII/\Ha\ ratio seen in figure \ref{fig3}, which improves the fit with the observations, and the decrease in the limiting  \OIII/\Hb\ ratio, which tends to make the fit somewhat worse at the low metallicity end.

Note that the aging tracks make a sharp zig-zag causing the 3~Myr isochrone to lie above the 2~Myr isochrone for metallicities above solar. This is particularly visible in panel (c) of figure \ref{fig3}, and  is the result of the appearance of the Wolf-Rayet stars at ages $\sim 3$~Myr, which briefly harden the overall stellar radiation field before they themselves explode as Type II supernovae.

\begin{figure*}
 \includegraphics[width=\hsize]{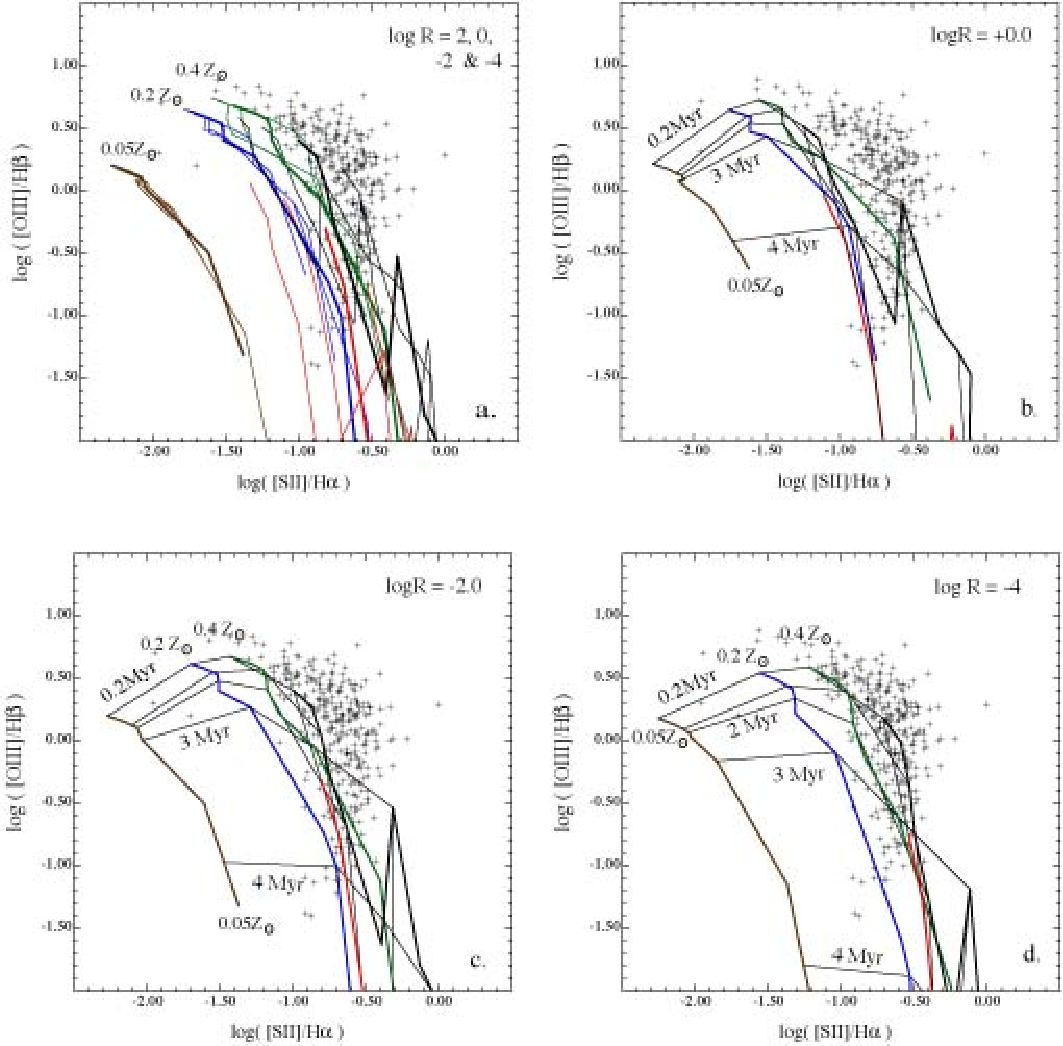}
  \caption{\label{fig4}
As figure \ref{fig3} but for the second of the V\&O diagnostics, the \SIILs /\Ha\ ratio \emph{vs.} the \OIII/\Hb\ ratio.}
\end{figure*}

The fit between theory and observation is much poorer when we consider the second of the Veilleux \& Osterbock diagrams which plots the \SIILs/\Ha\ ratio \emph{vs.} the \OIIIL/\Hb\ ratio. This is shown in figure \ref{fig4}. In general, the predicted  \SII\ lines are weaker than observed, by as much as a factor of two. A similar effect was seen in our earlier modeling, see figure 3 of \citet{Dopita00}, and is not fully understood. 

The  \SIILs/\Ha\ ratio is more ambiguous than the \NIIL/\Ha\ ratio if we wanted to use it as an abundance indicator, particularly at high metallicity. However, the 0.2 - 3~Myr isochrones still define a narrow range in line ratio space, albeit not exactly coincident with the observations. The reasons for such a discrepancy between theory and observation in this, as well as in other line ratio plots will be discussed in section \ref{discussion}.

\begin{figure*}
 \includegraphics[width=\hsize]{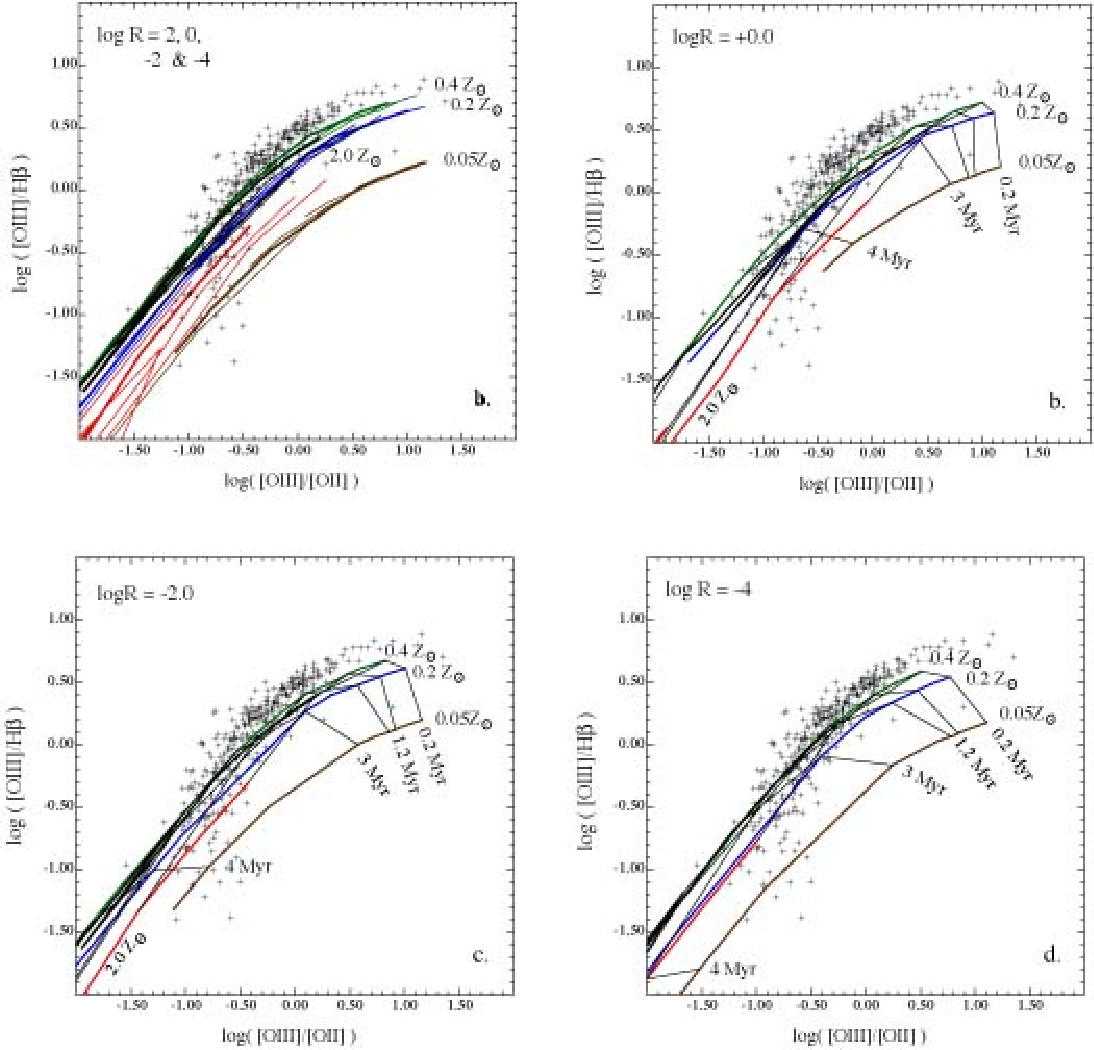}
  \caption{\label{fig5}
As figure \ref{fig3} but for the \OIIIL/\OIILs\ vs the \OIIIL/\Hb\ ratio. This is the diagnostic first used by \citet{Baldwin81} (BPT) to separate AGN from \HII\ regions.  High metallicity \HII\ regions ($ Z/Z_{\odot} > 2$) are located in the lower left hand corner of this diagram. Note also how how the low metallicity models fall systematically below the main sequence of observed \HII\ regions.}
\end{figure*}

\begin{figure*}
 \includegraphics[width=\hsize]{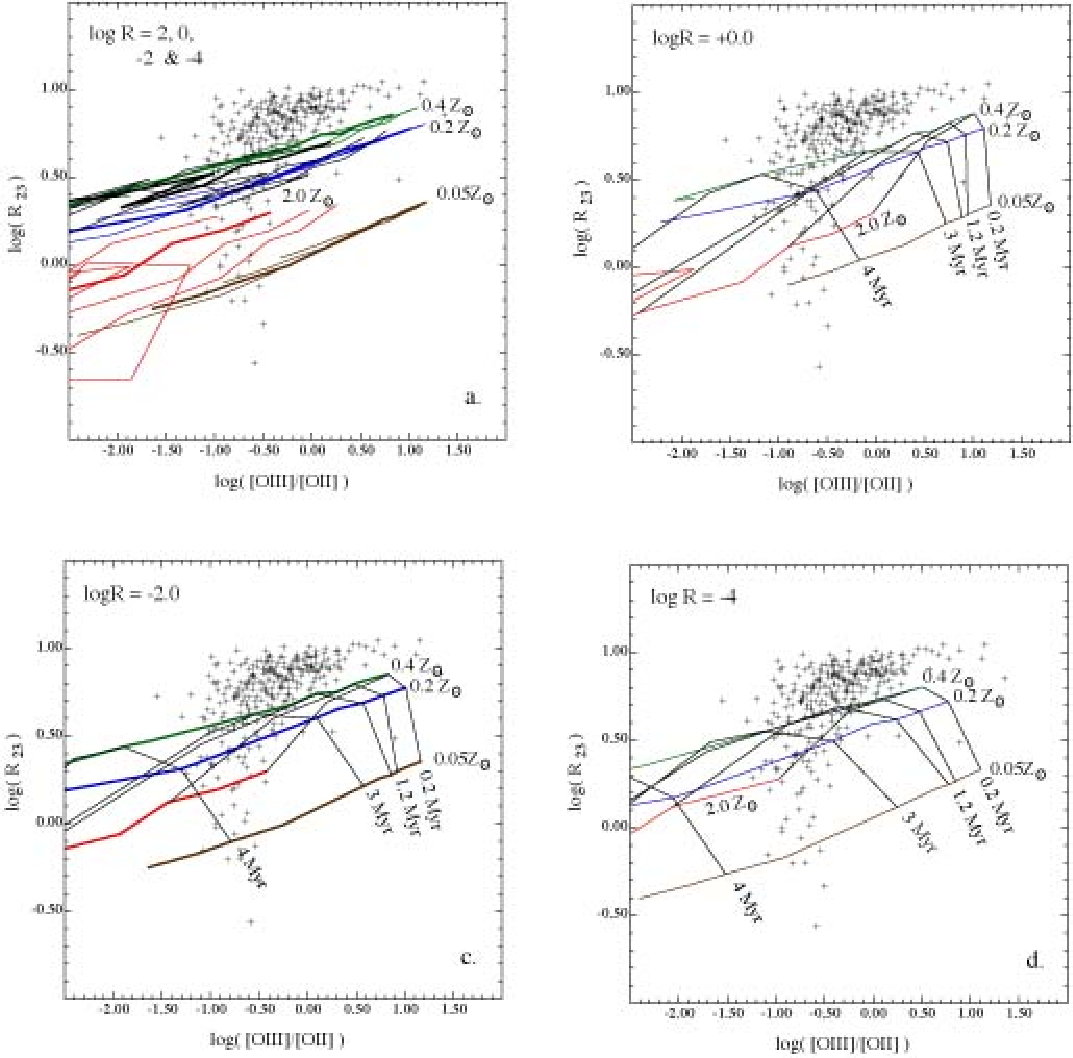}
  \caption{\label{fig6}
As figure \ref{fig3} but for the \R23| ratio \emph{vs.} the \OIIIL/\OIILs\ ratio. In panels (b) to (d) the solar metallicity track has been omitted for clarity. This shows that the traditional \R23\ ratio is not a good abundance diagnostic in the range $0.2 \leq Z/Z_{\odot} \leq 1.5$, approximately, and that it is quite sensitive to the nebular excitation. The theoretical tracks are, however, a rather poor fit to the observations. This is because the models have  \OIIILs\ and, even more so,  \OIILs\ weaker than those observed.}
\end{figure*}

\subsubsection{The BPT Diagram}
\citet{Baldwin81} (hereinafter BPT) first proposed the use of the \OIIIL/\OIILs\ ratio vs the \OIIIL/\Hb\ line ratio as a  diagnostic both to separate active nuclei from \HII\ regions and also to provide an excitation sequence for \HII\ regions. Our results for this line ratio are shown in figure \ref{fig5}. 
The \OIII/\OII\  ratio is an excitation sensitive ratio, sensitive to the ionization parameter in the nebula. Because of their higher ionization parameters and their central stars with higher effective temperatures at a given age, the low-metallicity \HII\ regions  are characterized by systematically higher \OIII/\OII\  ratios. Like all other ratios of optical forbidden lines to recombination lines, for a given stellar age, the \OIIIL/\Hb\ ratio starts off low at low metallicity, reaches a maximum at a certain metallicity, and then declines swiftly at higher metallicities. The reason for this was explained in the introduction in the context of the \R23\ ratio. As a consequence, the BPT diagram is highly degenerate in metallicity in the range  $0.2 \leq Z/Z_{\odot} \leq 2$, explaining the tightness of the observed \HII\ region sequence. This means, however, that the  \OIII/\Hb\ ratio is of little use as an abundance diagnostic except when $Z/Z_{\odot} \leq 0.2$. Even then it requires the observation of the \OIII/\OII\ ratio before an Oxygen abundance can be derived from it.

Although the theoretical sequence has much the same form as defined by the observations, it is notable that, for all values of  the \OIII/\OII\ ratio, the maximum of the \OIII/\Hb\ lies below the maximum defined by the observations. Furthermore, even for the highest value of $\log {\cal R}$, the zero-age models do not reach to the maximum \OIII/\OII\ ratio defined by the observations. As discussed in section \ref{discussion}, below, both of these discrepancies can be ascribed to a single source, namely that the input spectrum we are using in the models is  somewhat `softer' than the EUV spectrum found in real \HII\ regions.

\subsubsection{The \R23\ Ratio}
As described in the introduction, the \R23\ ratio has had a long and checkered history as a metallicity indicator. However, the sensitivity of the \OIIIL/\Hb\ ratio to the cluster age shown in figure \ref{fig4} suggests that we should plot the \R23\ ratio against  the \OIII/\OII\  ratio in order to be able to investigate the age sensitivity of the \R23\ ratio. This diagnostic is shown in figure \ref{fig5}.

For this diagnostic the fit between theory and observation is quite poor, due mainly to the fact that the \R23\ ratio is systematically underestimated in the models. The effect is larger than that seen in the  \OIII/\Hb\ ratio (see figure \ref{fig5}), showing that predicted strengths of the \OIII\ but especially  the \OII\ lines weaker than observations suggest. This will be discussed in the Discussion section, below. Points with low \R23\ would be explained as having metallicities greater than twice solar. 

This diagnostic plot shows immediately why the calibration of the \R23\ ratio in terms of metallicity has been so difficult. Not only is the range of \R23\ rather restricted, but is it degenerate in terms of metallicity over a rather generous range,  $0.2 \leq Z/Z_{\odot} \leq 1.5$, approximately. Not only that, but the ratio is fairly sensitive to excitation, as measured by the \OIII/\OII\  ratio. Because low metallicity \HII\ regions are characterized by high ionization parameters, the low metallicity points tend to occur in the upper right of this plot, while, from the form of the isochrones at high metallicity, the spray of points towards the bottom left of this diagnostic can only be due to very high metallicity ($ Z/Z_{\odot} > 2$) \HII\ regions with ages of 1--3~Myr. 

Based on these problems, we are constrained to conclude that \R23\ is not a good abundance diagnostic. We will return to this point in Section \ref{discussion}, below.

\subsubsection{\citet{Dopita00} Diagnostics}

\citet{Dopita00} introduced two new abundance diagnostics, based on the \NII/\OII\ line ratio. This ratio was favoured for two reasons which make it  {\em very} sensitive to abundance. The first is that, over much of the range of interest, nitrogen is a secondary nucleosynthesis element, or at least has a large secondary component. Thus the ratio  \NII/\OII\ will systematically increase with abundance. The second is connected with the fact that the \OIILs\ lines are found in the near UV part of the spectrum, whereas the \NIIL\ is in the red part of the spectrum. This might be construed as a disadvantage in the use of the \NII/\OII\ ratio, since the reddening corrections are large, and correspondingly more uncertain. However, this problem is more than offset by the fact that the mean temperatures of \HII\ regions are a strongly decreasing function of metallicity  and at high abundances, the mean thermal energy of the electrons becomes too low to excite transitions with large electron excitation energy. Thus transitions such as the \OIILs\ lines are quenched, while the lower threshold energy \NIIL\ continue to be excited. For this reason as well, the  \NII/\OII\ line ratio falls with increasing abundance.

\begin{figure*}
 \includegraphics[width=\hsize]{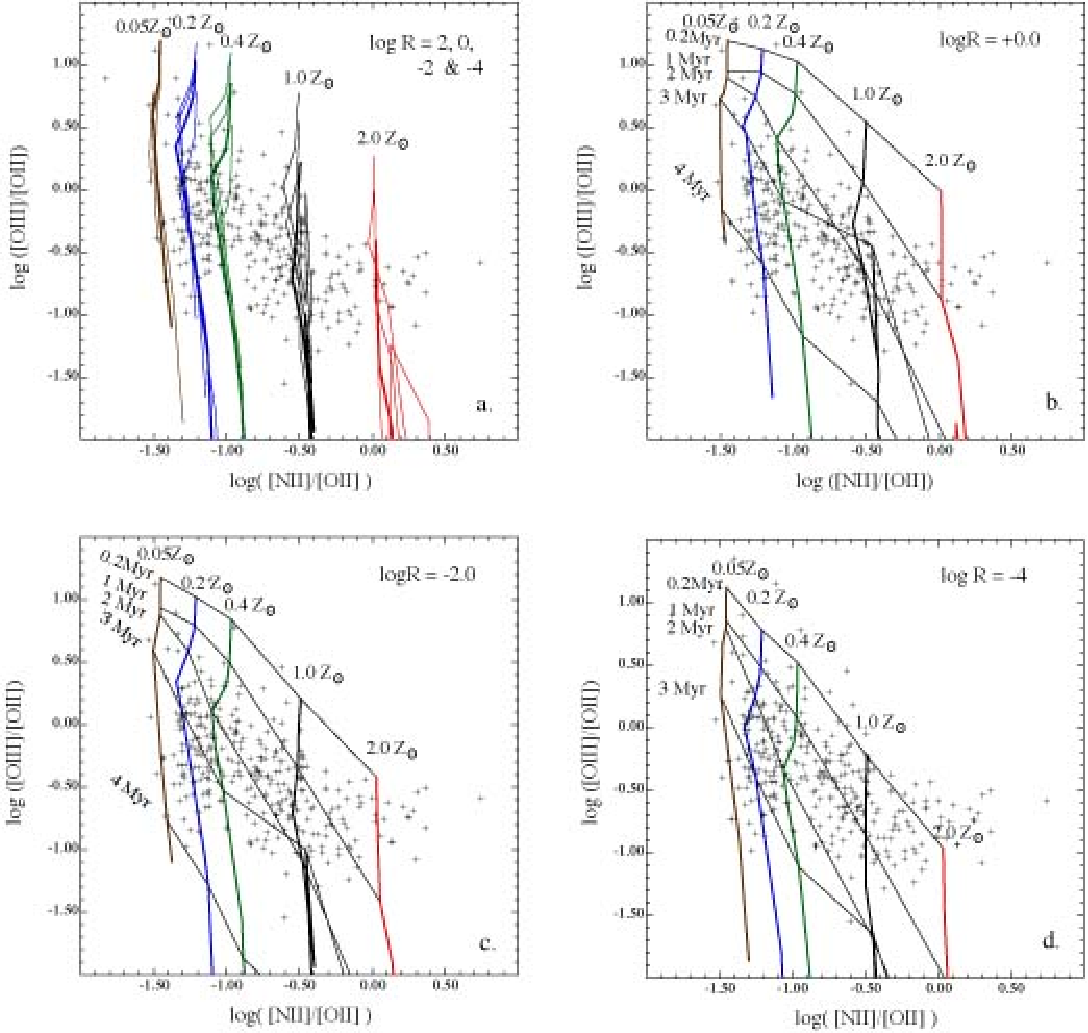}
\caption{\label{fig7}
As figure \ref{fig3} but for the \NIIL/\OIILs\ against the \OIIIL/\OIILs\ ratio used by \citet{Dopita00}. This is an excellent diagnostic of both metallicity and of stellar age. Note that the  \OIIIL/\OIILs\ ratio is sensitive to \emph{both} age and metallicity. The relationship between the observational points and the isochrones in this diagram suggests that most \HII\ regions are observed in the age range 1--3~Myr.}
\end{figure*}

The systematic decrease in the electron temperature with increasing metallicity may also affect the \OIIIL/\OIILs\ ratio. By itself, this would induce an increase in the \OIII/\OII\ ratio with metallicity. However, this is more than offset by two factors. Firstly, the mean effective temperature of the exciting stars falls with metallicity, and secondly, the mean ionization parameter is itself a decreasing function of metallicity. Both of these lead to a net decrease in the  \OIII/\OII\ line ratio with metallicity. Nonetheless, the  \OIII/\OII\  line ratio remains a sensitive indicator of the ionization parameter, which is strongly age dependent. Therefore, a plot of \NIIL/\OIILs\ against the \OIIIL/\OIILs\ ratio should separate stellar age from stellar abundance. This is illustrated in figure \ref{fig7}.

It is clear that the \OIIIL/\OIILs\ ratio is an excellent diagnostic of both the stellar age and the nebular abundance. According to the theoretical tracks, all the observed \HII\ regions have ages between 0 and roughly 4~Myr, with most lying in the age range 1--3~Myr. The division into abundance bins by use of the  \NIIL/\OIILs\  is very clean. Provided that the calibration is correct, then the abundance can be obtained using only the  \NII/\OII\  ratio to a precision of roughly 0.15~dex.

\begin{figure*}
 \includegraphics[width=\hsize]{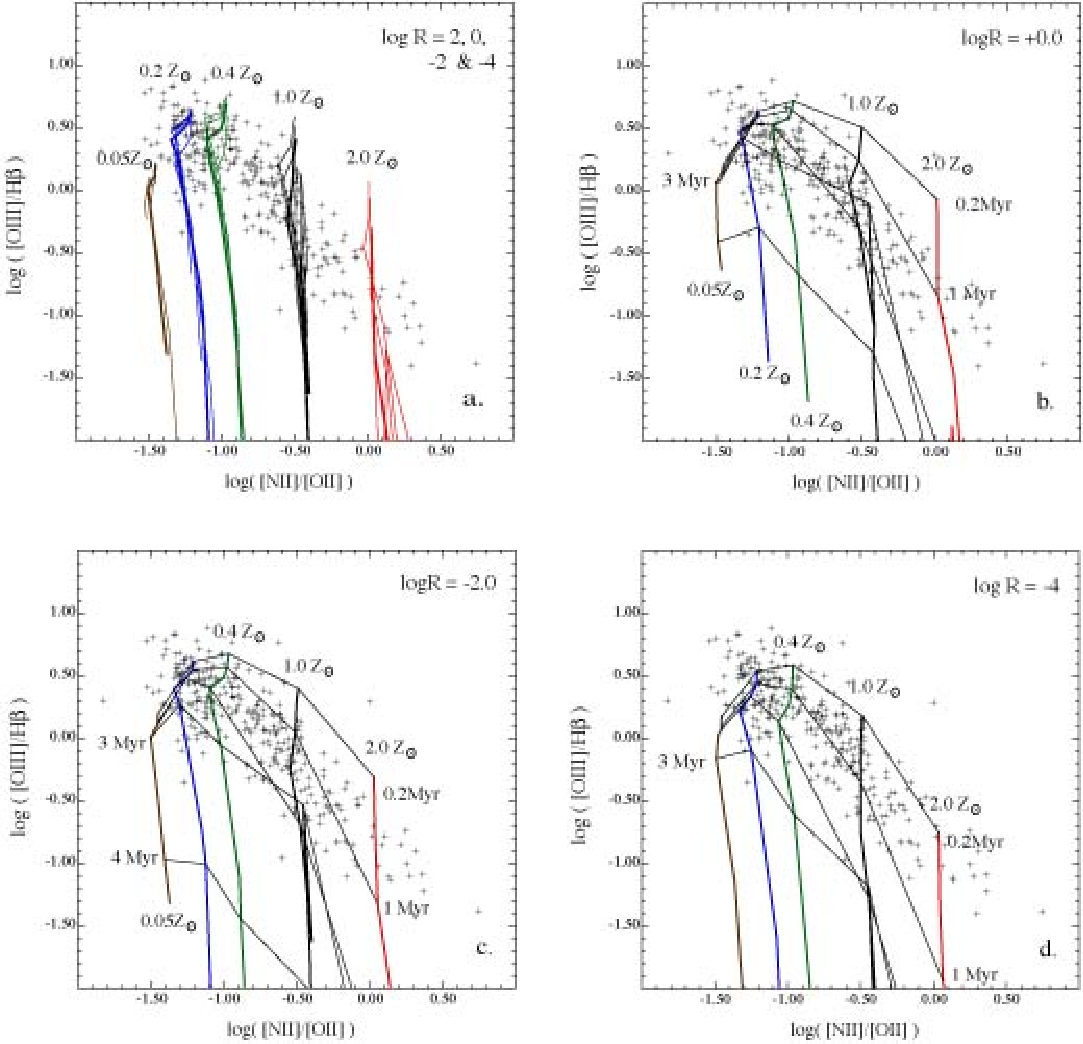}
\caption{\label{fig8}
As figure \ref{fig3} but for the \NIIL/\OIILs\ against the \OIIIL/\Hb\ ratio, as used by \citet{Dopita00}. This is also a good diagnostic of both metallicity and of stellar age. Given that the models tend to underestimate the \OIIIL/\Hb\ ratio, the age of the \HII\ regions using this diagnostic is also likely to be an underestimate.}
\end{figure*}

The \OIIIL/\Hb\ line ratio can be substituted for the \OIII/\OII\ ratio, since both are sensitive to ionization parameter and therefore to the age of the cluster. However, the separation into age classes is not so sharp using this ratio, as can be seen in figure \ref{fig8}. Furthermore, we have seen in figure \ref{fig5} that the models tend to systematically underestimate the \OIIIL/\Hb\ ratio, so therefore it follows that the age of the \HII\ regions inferred using this diagnostic is also likely to be an underestimate of the true age.

\begin{figure*}
 \includegraphics[width=\hsize]{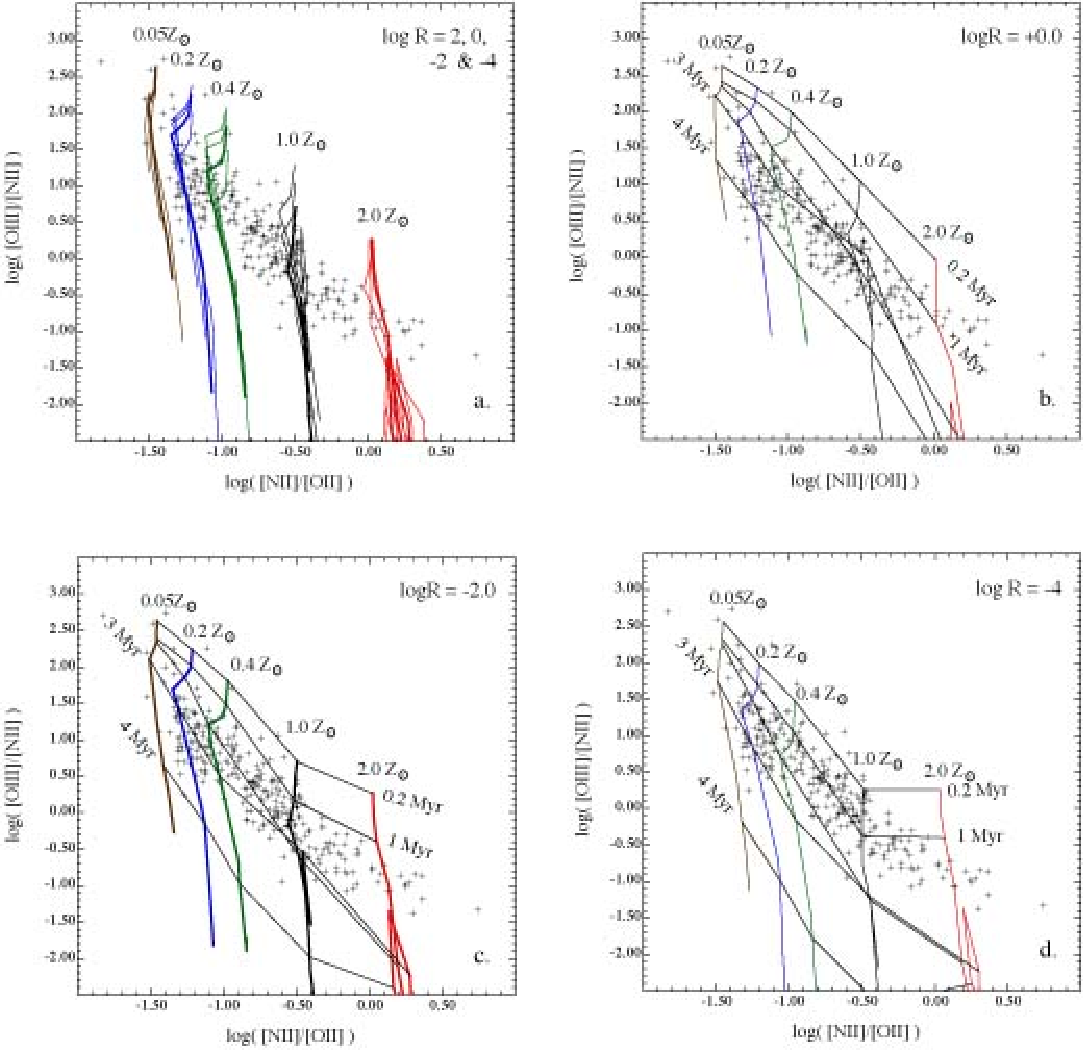}
\caption{\label{fig9}
As figure \ref{fig3} but for the \NIIL/\OIILs\ against the abundance-sensitive  \OIIIL/ \NIIL\ ratio, as proposed by \citet{Alloin79}. Although the \OIII/\NII\ ratio is indeed sensitive to abundance, it is also very sensitive to ionization parameter as shown by \citet{Dopita00} and therefore to stellar age. This diagram suggests that most \HII\ regions are observed in the age range 1--3~Myr, in good agreement with age estimate obtained by figure \ref{fig7}.}
\end{figure*}

The third  \citet{Dopita00} diagnostic which utilizes the  \NIIL/\OIILs\  ratio is that which plots this ratio against the  \OIIIL/ \NIIL\ ratio. The  \OIIILs/ \NIILs\ ratio was first proposed by \citet{Alloin79} as a more convenient abundance diagnostic, which is similar to the one we are using here.  The observations and theoretical curves on this diagnostic are shown in figure \ref{fig9}. As can be seen in the diagram, both axes are quite sensitive to abundance, but the  \citet{Alloin79} ratio, by its use of the \OIIIL\ in the place of the \OIILs\ is also more sensitive to parameters other than the abundance.

\cite{Pettini04} have argued that the \OIIIL/ \NIIL\ ratio is of itself a sensitive abundance indicator suitable to be used in the analysis of abundances in high-redshift galaxies. This is a little too simplistic. Because the  \OIII/\NII\ ratio is strongly dependent upon the excitation of the nebula, it is also sensitive to both the ionization parameter, as shown by \citet{Dopita00} and to the age of the cluster of exciting stars. These sensitivities are as great as the sensitivity to abundance alone. Nonetheless, the \NIIL/\OIILs\ \emph{vs.}  \OIIIL/ \NIIL\ diagnostic plot has great utility in separating abundance from cluster age, and the results that it gives are in broad agreement with figure \ref{fig7}. From these two plots we can conclude that most \HII\ regions are observed in the age range 1--3~Myr with little sensitivity of this age range with abundance. 

This is in full accord with expectations. As shown in Paper II, after an age of 3~Myr, the absolute luminosity of \HII\ regions in the recombination lines drops sharply, and due to the expansion of the \HII\ region, the surface brightness drops even more rapidly. Both of these militate strongly against the observation of old \HII\ regions in extragalactic studies. Furthermore, very young \HII\ regions tend to be dust obscured since they are still embedded in their placental molecular cloud complex. Such \HII\ regions will not be bright at optical wavelengths. In Paper II we estimated the molecular cloud dissipation timescale for the molecular clouds to be about 1~Myr using the particular example of M17. The molecular cloud dissipation timescale was introduced in Paper I, and is the characteristic timescale $\tau$ for the \HII\ region to emerge from its compact or ultra-compact configuration, and is a measure of the solid angle (as measured at the exciting stars) which is intercepted by molecular clouds; $\Omega(t)/4\pi = \exp[-t/\tau]$. Clearly, the relative dearth of young \HII\ regions beween 0.2 and 1.0~Myr, evident in figures \ref{fig7} and \ref{fig9}, is consistent with $\tau \sim 1$~Myr.

In conclusion, the \NIIL/\OIILs\  ratio \emph{vs.} the  \OIIIL/ \NIIL\ provides a good diagnostic of both metallicity and age of the exciting cluster. The \OIIIL/ \NIIL\  ratio can only be used by itself as an abundance indicator to the degree to which we can assume a (fairly restricted) characteristic age for the exciting clusters. Statistically, this may be valid, but the results of its use in single galaxies should be treated with caution.

\subsubsection{\NII/\Ha\ vs \SII/\Ha}
\begin{figure*}
 \includegraphics[width=\hsize]{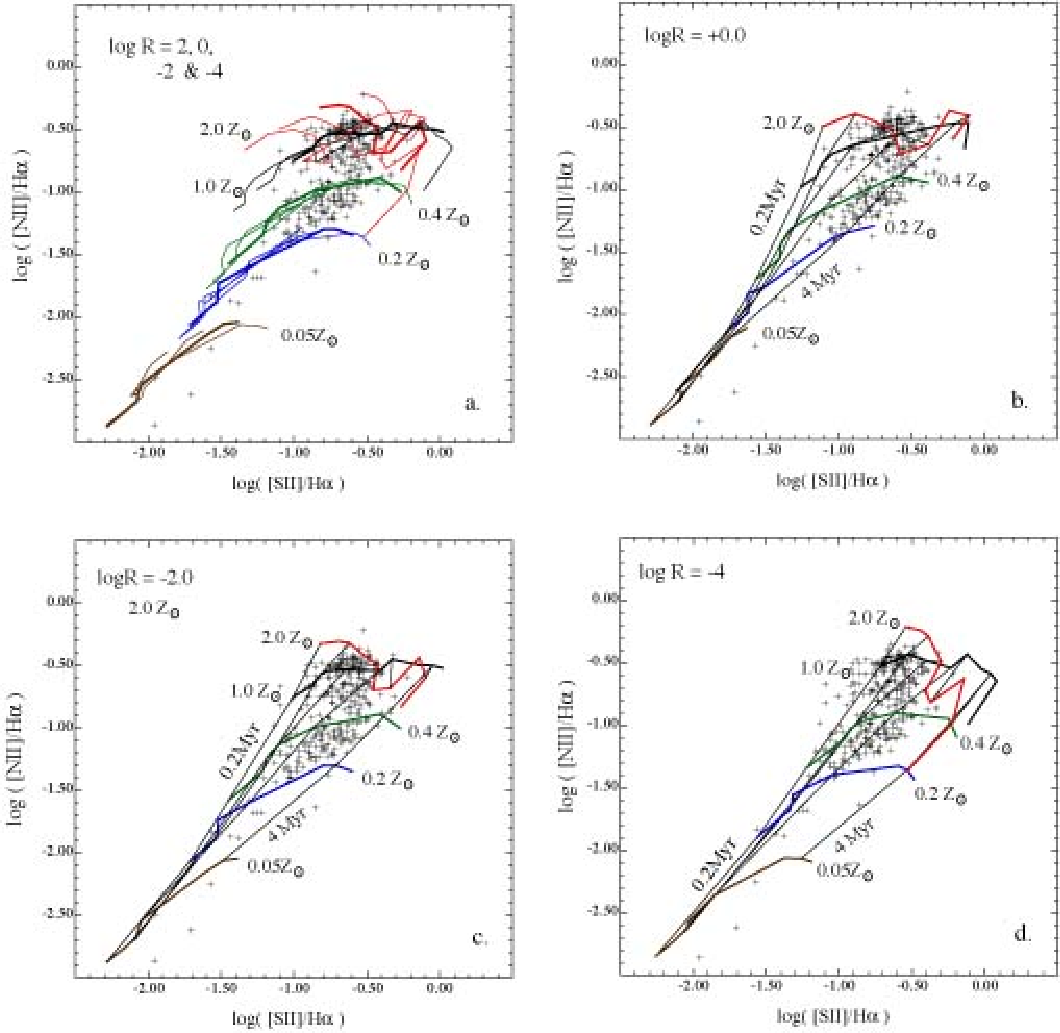}
\caption{\label{fig10}
As figure \ref{fig3} but for the \NIIL/\Ha\ \emph{vs.} \SIILs/\Ha\ diagnostic line ratios. As in Figure \ref{fig3} the data for individual \HII\ regions are shown as crosses, and the integrated spectra from \citet{Moustakas05} are shown as points. Because these are both ratios of lines in the red portion of the spectrum, and because N is partially a secondary element, this diagnostic turns out to be a good abundance diagnostic for $Z/Z_{\odot} < 1.0$. The models of the older and more diffuse \HII\ regions are located close to the right hand boundary of the observational points. This gas can be identified with the warm ionized medium (WIM) in galaxies.}
\end{figure*}

Because they lie very close  to each other in wavelength, and because we have seen that \NII\ is very sensitive to abundance, the \NIIL/\Ha\ \emph{vs.} \SIILs/\Ha\ plot offers some hope as an abundance diagnostic. This has recently been investigated by \citet{Moustakas05}, who were interested to find out whether  integrated spectra of galaxies were different from single \HII\ regions when plotted in this way. We would expect to find such a difference in principle, because the integrated spectra do not discriminate against the emission from old \HII\ regions in the manner described in the previous section. The results of our modelling is shown in figure \ref{fig10}.

In Paper II, we found that the diffuse component due to old, evolved, large and faint \HII\ regions could amount to 20--30\% of the total \Ha\ emission, and suggested that this could be identified with the Diffuse Ionized Medium (DIM) which is almost ubiquitously seen in disk galaxies \citep{Hoopes96,Martin97,Wang98, Wang99}. This is usually refereed to as the ``Reynolds Layer" in the context of our own Galaxy; see a recent review by \citet{Reynolds04}.  In the past, the  excitation of the DIM has been usually ascribed to leakage of ionizing photons from young \HII\ regions. However, the emission from large, faint, low pressure and evolved \HII\ regions might provide a better explanation.

In figure \ref{fig10}, the models of old \HII\ regions with metallicities $Z/Z_{\odot} > 1.0$ reach up to line ratios of \SIILs/\Ha\ $\sim 1$ and \NIILs/\Ha\ $\sim 2/3$, which is certainly consistent with observed line ratios encountered in the DIM. However, in our Galaxy, line ratios even greater than these are seen. To explain these, we would have to invoke an additional heating mechanism, such as shocks, or a harder EUV field possibly produced by old ex-nuclei of Planetary Nebulae, which are sufficiently numerous to produce an appreciable EUV field out of the plane of the Galaxy.

\subsection{Infrared Line Diagnostics}

The ISO satellite first demonstrated the utility of mid-IR line diagnostics in the analysis of \HII\ region spectra. For normal galactic \HII\ regions,  \citet{Giveon02} obtained extensive ISO spectroscopy, as did \citet{Peeters02} in the context of Ultra-Compact (UC) \HII\ regions. Later, \citet{Verma03} measured integrated spectra for a number of famous Starburst galaxies. We will use these data here, although we note that both the high-resolution and low-resolution modes of the Infrared Spectrometer (IRS) \citep{Houck04} on the   \emph{Spitzer Space Observatory} are producing much new and excellent data on both individual \HII\ regions as well as on spectroscopy of whole galaxies.

\begin{figure}
 \includegraphics[width=\hsize]{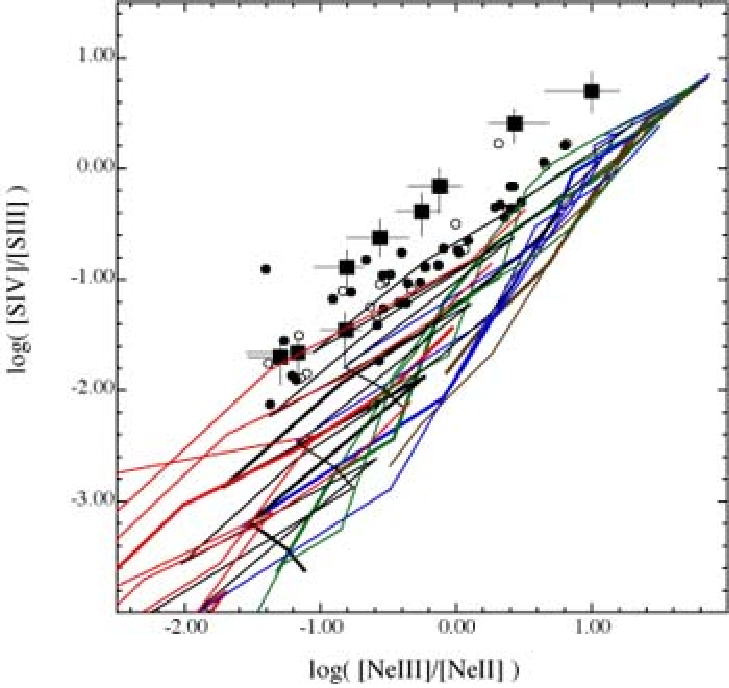}
\caption{\label{fig11}
 The [\ion{Ne}{3}] 15.5\mum\ /  [\ion{Ne}{2}] 12.8\mum\ ratio versus the [\ion{S}{4}] 10.5\mum\ /  [\ion{S}{3}] 18.7\mum\ ratio. \HII\ regions from \citet{Giveon02} are plotted as filled circles, UC \HII\ from \citet{Peeters02} as open circles, and Starburst galaxies from \citet{Verma03} as filled squares with error bars. The theoretical curves create as superb an example of theoretical spaghetti as anything these authors have seen.}
\end{figure}

Within the spectral range of the IRS, such line ratios that can be constructed are primarily sensitive to the nebular excitation. Examples include the [\ion{Ne}{3}] 15.5\mum\ /  [\ion{Ne}{2}] 12.8\mum\ ratio,  the  [\ion{S}{4}] 10.5\mum\ /  [\ion{S}{3}] 18.7\mum\ ratio, the [\ion{S}{4}] 10.5\mum\ / [\ion{Ar}{3}] 9.0\mum\ ratio, and the [\ion{Ne}{3}] 15.5\mum\ / [\ion{S}{3}] 18.7\mum\ ratio. Unfortunately, these are of little use in measuring anything other than the hardness of the EUV radiation field. All of the strong lines accessible in this region of the spectrum are produced by $\alpha -$process elements, and therefore the abundance ratios stay approximately constant with abundance. All the lines have low energy thresholds for excitation, and therefore none of them are particularly sensitive to the electron temperature, which varies strongly with abundance.  In figure \ref{fig11} we present only one example to demonstrate how poor these line ratios are as abundance diagnostics when used alone. 

Much of the data used in this plot refers to observations of ultra-compact \HII\ regions, which are not exactly comparable to ``normal" \HII\ regions. In particular, they are extremely dense, most are excited by only a single OB star, and many of the ionizing photons from the exiting star are absorbed by dust rather than by the ionized plasma. This accounts for much of the offset between the (low density) models and the data in figure \ref{fig11}. This effect is particularly important for the  [\ion{S}{4}] 10.5\mum\ /  [\ion{S}{3}] 18.7\mum\ ratio. These points, and this particular diagnostic plot was thoughly discussed in our paper on compact \HII\ regions \citep{Dopita06b}.

\begin{figure}
 \includegraphics[width=\hsize]{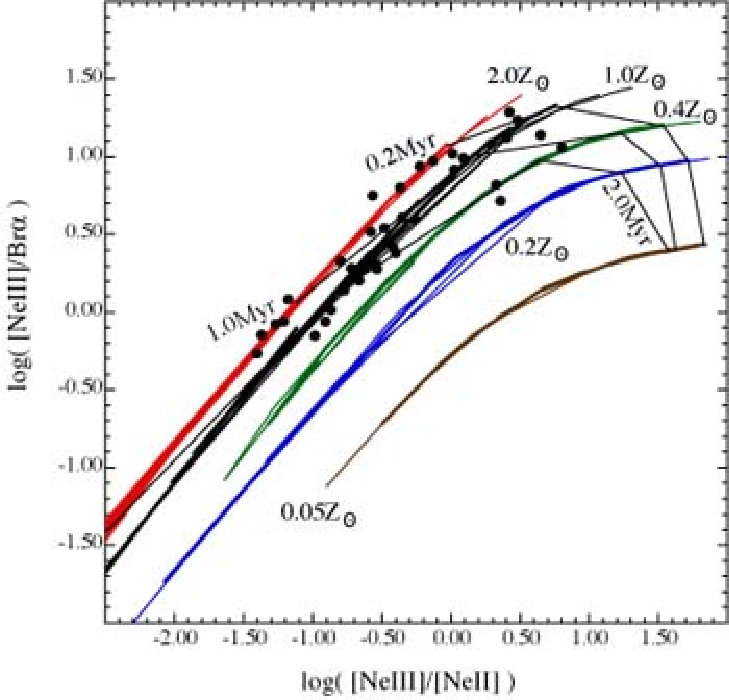}
\caption{\label{fig12}
 The [\ion{Ne}{3}] 15.5\mum\ /  [\ion{Ne}{2}] 12.8\mum\ ratio versus the [\ion{Ne}{3}] 15.5\mum\ /  Br$\alpha\ \lambda 4.051$\mum\ line ratio. \HII\ regions from \citet{Giveon02} are plotted as filled circles. This diagnostic provides a clean abundance determination which is monotonic over the full abundance range. The isochrones are given for the case $\log {\cal R}=-2$. This diagram is an infrared analog of fig \ref{fig5}, but is a much cleaner diagnostic because these IR line ratios are very insensitive to the nebular temperature.}
\end{figure}

In order to form a line ratio that is sensitive to abundance, we need to use a ratio with a Hydrogen or Helium recombination line. Unfortunately, these are rather faint in the mid-IR. About the best line we have is Br$\alpha\ \lambda 4.051$\mum. Alternatively, Pfund$-\alpha\ \lambda 7.458$\mum\ might be useable, although it is often difficult to observe against the PAH features seen at this wavelength. In figure \ref{fig12} we show one such diagnostic, the [\ion{Ne}{3}] 15.5\mum\ /  [\ion{Ne}{2}] 12.8\mum\ ratio versus the [\ion{Ne}{3}] 15.5\mum\ /  Br$\alpha\ \lambda 4.051$\mum\ line ratio. This has the advantage that, at a given value of the excitation measured by the  [\ion{Ne}{3}] 15.5\mum\ /  [\ion{Ne}{2}] 12.8\mum\ ratio, the [\ion{Ne}{3}] 15.5\mum\ /  Br$\alpha\ \lambda 4.051$\mum\ line ratio is a monatonic function of abundance up to at least $Z \sim 2.0\times Z_{odot}$, and so it provides a clean abundance diagnostic, unlike the comparable optical line ratio shown in figure \ref{fig5}.

Currently, the uncertainties in the observational data are to large for this to provide a very accurate abundance diagnostic. However, we may note that the observations from \citet{Giveon02} are consistent with abundances between 2.0 and 0.4 $Z_{\odot}$, and with ages between 0 and 2~Myr. This abundance range is exactly what we would expect to find for Galactic \HII\ regions over the range of Galactocentric radii observed by him. It is interesting that the ages are lower than those derived for  \HII\ regions at optical wavelengths. This is consistent with the selection criteria for IR-bright, compact \HII\ regions. This selects for young \HII\ regions. Additionally, the use of mid-IR wavelengths allow the \HII\ region emission to be seen almost unimpeded by the surrounding molecular material, so the molecular cloud dissipation timescale is not a relevant parameter here.

\section{Integrated Spectra}\label{mean}

The diagnostics presented in the previous sections were designed to derive parameters for single \HII\ regions. However, what we seek to do now is to develop a spectral synthesis by taking recombination line flux averaged spectra along an evolutionary track of an \HII\ region with given parameters. This  reduces the number of free parameters to two; $Z/Z_{\odot}$ and $\log {\cal R}$. The observations can then be used to constrain the values $\log {\cal R}$, allowing the extragalactic abundance sequence for ensemble averages of evolving \HII\ regions to be derived. This technique opens the way to the accurate study of the metallicity evolution of both individual galaxies and the Universe.

The model we adopt assumes that clusters are formed continuously, and that any single cluster has all of its stars born coevally. As described above, the cluster mass function does not enter explicitly into the models, which are characterized by their $R-$values. The process of producing an integrated spectrum for any $Z$ is therefore simply to integrate the fluxes from a single \HII\ model through time, until it has faded such that $\ge 96$\% of all ionizing photons have been emitted. This limits the integration time to lie within the first $5-6.5$~Myr. Stochastic effects due finite sampling of the stellar IMF - important at low cluster masses - are not included. However, the form of the cluster initial mass function either as inferred in Paper II, or derived from stellar number counts is such that each logarithmic bin of cluster mass contributes a similar amount to the total flux. Massive clusters will therefore tend to wash out the stochastic effects of low-mass clusters.

On line ratio diagnostics, there is no simple relationship between the ensemble averaged line ratio and the line ratios in the individual \HII\ regions which make up this average. For example, a high ionization line such as \OIIIL\ will be mostly produced in the very youngest \HII\ regions with the highest $R-$values. Low ionization lines such as the \SIILs\ lines will have a much greater weighting from older evolved \HII\ regions with low $R-$values. Indeed, for this line the majority of the line flux may come from regions lying outside the bright \HII\ regions that would be normally selected for observation in studies of individual \HII\ regions - in other words, from regions that form part of the `warm ionized medium' (WIM).

Given these quite different selection biases, it is interesting to compare line ratio observations of the integrated line fluxes of galaxies with those of individual \HII\ regions. Such a comparison helps us to quantify the importance of the WIM in determining the global line ratio. For this purpose, we have used the observations of individual \HII\ regions listed and used above ( \ref{fig3} and references in Section 3.1), and we have drawn our sample of integrated line fluxes of galaxies from the Nearby Field Galaxy Survey (NFGS) \citep{Jansen00}, and the recent survey by \citet{Moustakas05}.

\begin{figure}
 \includegraphics[width=\hsize]{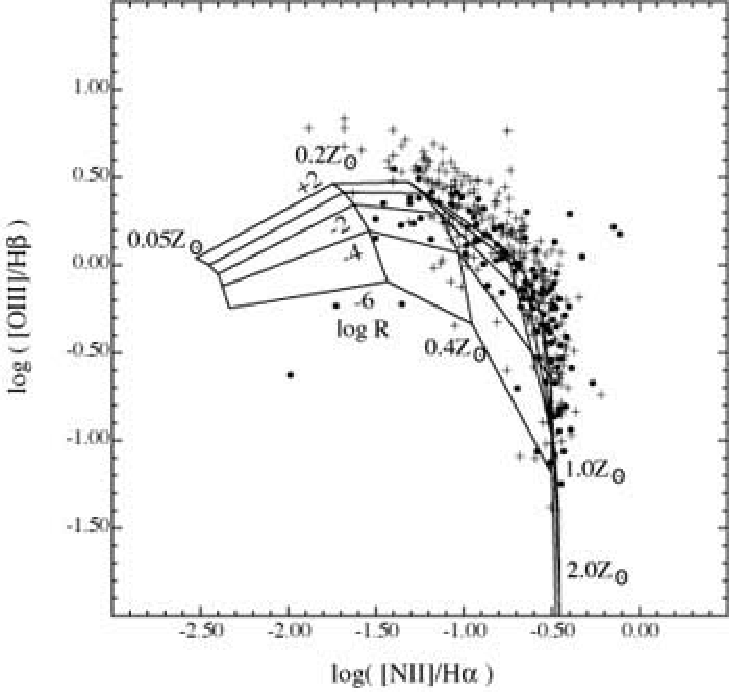}
\caption{\label{fig13}
 The synthesised \HII\ region spectra on the V\&O diagnostic \NIIL/\Ha\ \emph{vs.} \OIIIL/\Hb\  (c.f. figure \ref{fig3}). The integrated galaxy spectra from the Nearby Field Galaxy Survey (NFGS) \citep{Jansen00} are shown as filled circles, while the crosses are for individual \HII\ regions drawn from the sources referenced above, to show the degree to which the integrated spectra differ from the single source spectra. Note the large disparity between the integrated spectra and the observations of individual \HII\ regions at the low abundance end ($Z/Z_{\odot} < 0.4$). Compared with earlier models, the theory provides a good fit to the integrated galaxy data providing that $\log {\cal R} > -4$, and emphasizes the degeneracy of this V\&O diagnostic. }
\end{figure}

In figure \ref{fig13} we plot the main V\&O diagnostic \NIIL/\Ha\ \emph{vs.} \OIIIL/\Hb. This should be compared with figure \ref{fig3}. The models provide a good fit to the integrated spectra of galaxies. Indeeed, the fit is better than that of the models  of individual \HII\ regions compared to data in figure \ref{fig3}. This is especially apparent at the low abundance end (where most observations are of relatively nearby dwarf irregular galaxies such as NGC6822, IC1613, IIZw40 and the like). This suggests that the observations of these low-abundance \HII\ regions may be affected by aperture effects. This would be caused by observers picking out bright high-excitation sub-regions of the \HII\ region from their long-slit spectra for analysis, rather than observing across the whole \HII\ region to obtain an integrated spectrum of it. This would preferentially enhance the \OIIIL/\Hb\ ratio while at the same time depressing the  \NIIL/\Ha\ ratio, moving the observed point in line ratio space towards the top and the left.

\begin{figure}
 \includegraphics[width=\hsize]{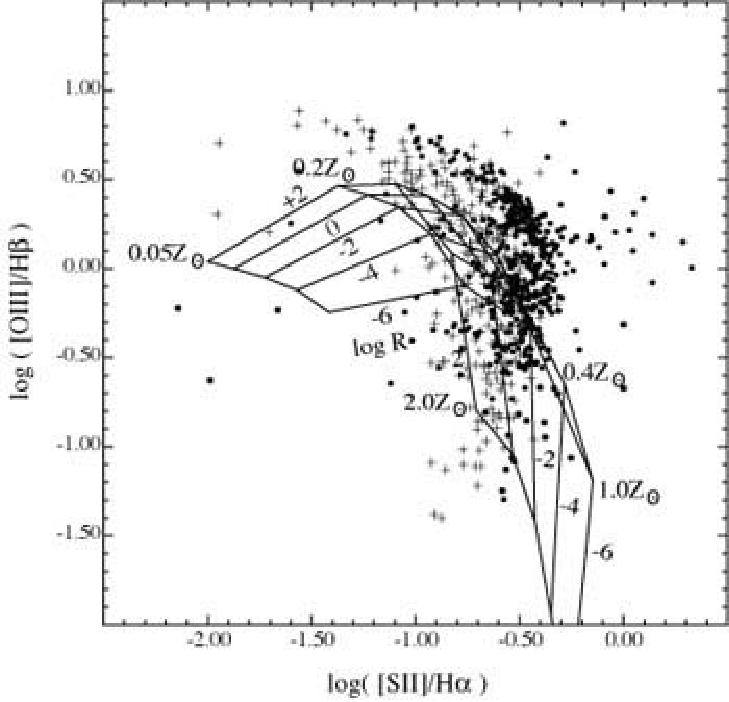}
\caption{\label{fig14}
 The synthesised \HII\ region spectra on the V\&O diagnostic   \SIILs/\Ha\ \emph{vs.}  \OIIIL/\Hb\  (c.f. figure \ref{fig5}). The integrated galaxy spectra are shown as filled circles, and include both the NFGS data and the data from \citet{Moustakas05}. Both the  \SIILs\ and the  \OIIIL\ lines seem to be a little weak in the models compared with the observations.}
\end{figure}

The fit with the second V\&O diagnostic, the  \SIILs/\Ha\ \emph{vs.} \OIIIL/\Hb\  (figure \ref{fig14}) is less convincing. The integrated spectra are displaced to the right on the \SIILs/\Ha\  ratio compared with the spectra from individual \HII\ regions, showing the importance of the contribution of the more diffuse ionized component. It is interesting to note that most of the highest points in terms of the \OIIIL/\Hb\  ratio are contributed by the  \citet{Moustakas05} data set. In this respect there are systematic differences between this data set and the NFGS survey. The reason for this is not apparent to us.

In the models, the \NIILs\ are the major coolants in the outer parts of the \HII\ regions. It is possible that the abundance of Nitrogen in the ionized gas is somewhat overestimated (by an underestimate of the depletion, for example), leading to saturation in the \NIIL/\Ha\ ratio, and a reduction in the \SIILs/\Ha\  ratio.
The problem with the models in reproducing the observed \SIILs/\Ha\ ratios is systematic. Even with the old \emph{Starburst 99 v2} models, a similar problem was encountered by \citet{Dopita00}. This will be discussed further in the next section.

\begin{figure}
 \includegraphics[width=\hsize]{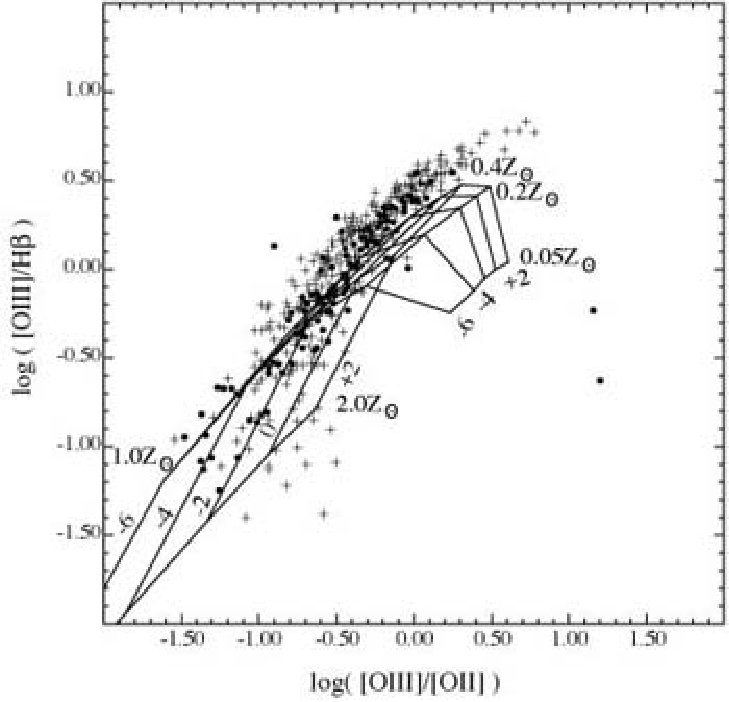}
\caption{\label{fig15}
 As figure \ref{fig13}, but for the BPT diagram, the \OIIIL/\OIILs\ ratio vs \OIIIL/\Hb\ . This should be compared with figure \ref{fig5}.}
\end{figure}

Let us now turn our attention to the BPT diagram, the \OIIIL/\OIILs\ ratio \emph{vs.} the \OIIIL/\Hb\ ratio \citep{Baldwin81}. This is plotted in figure \ref{fig15}. Once again, the fit between the theory and observation is good, although on this diagnostic the data once again fall somewhat high relative to the theory in the \OIIIL/\Hb\ ratio at the low abundance end $Z/Z_{\odot} < 0.4$. 

We note that the \NIIL/\Ha\ ratio covers quite a large range and displays considerable sensitivity to abundance. However, the same is not true of the  \OIIIL/\Hb\ ratio. Only two points in the NFGS survey lie below $\log$\NIIL/\Ha\ $= -1.5$, and since the   $Z/Z_{\odot}$ curve lies to the left of $\log$\NIIL/\Ha\ $= -1.5$, figure \ref{fig13} suggests that the lower abundance bound to the NFGS survey is about $Z/Z_{\odot} < 0.2$. This is also consistent with what would be implied from figure \ref{fig15}. Both figure \ref{fig14} and figure \ref{fig15} suggest that the upper bound is about $Z/Z_{\odot} \sim 2$.

\begin{figure}
 \includegraphics[width=\hsize]{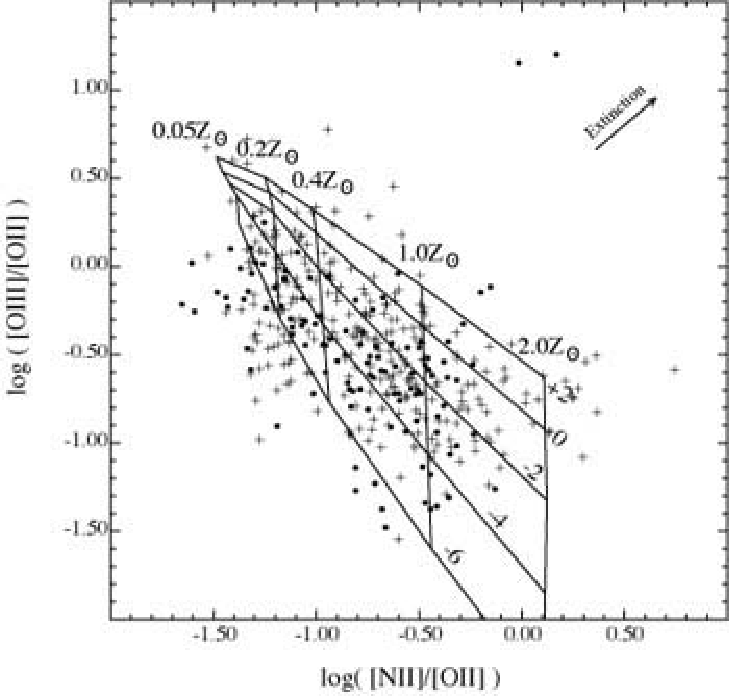}
\caption{\label{fig16}
 As figure \ref{fig13} but for the \NIIL/\OIILs\ ratio against the \OIIIL/\OIILs\ ratio used by \citet{Dopita00}. This should be compared with figure \ref{fig7}. This is perhaps the most unambiguous diagnostic for the determination of both metallicity and $\log {\cal R}$. For the meaning of the arrow marked `extinction' see the text.}
\end{figure}

The first of the \citet{Dopita00} diagnostics is shown in figure \ref{fig16}. This plots the \NIIL/\OIILs\ ratio against the \OIIIL/\OIILs\ ratio, and should be compared with figure \ref{fig7}. There is very little distinction between the distribution of the integrated spectra from the NFGS survey and the observations of individual \HII\ regions on this diagram. Essentially any value of $\log {\cal R}$ seems to be permitted, although the majority of the points are consistent with a range 0 to -4 in this parameter. This is consistent with what we might have expected,  given that the parameters which enter into $R$ , $M_{\rm cl}$ and $P/k$, cover at least the range $ 100 < M_{\rm cl}/M_{\odot} < 10^5$ and $10^4 < P/k < 10^7$cm$^{-3}$K.

The abundance range of the NFGS survey suggested by this plot is broader than that given by the V\&O or BPT diagnostics, since the theory implies the existence of some galaxies with very low abundances. However, we must beware once again of the observational problems which enter into plots like this. The \NIIL\ line and the \OIILs\ lines lie at opposite ends of the spectrum, and so there is a very large reddening correction applied to the data. Generally speaking a standard reddening law is used to do this. In the case of the NFGS survey, for example the \citet{Clayton89} extinction curve was used. However, the foreground dust screen in front of an individual \HII\ region (let alone a whole galaxy) has a complex fractal structure, and rather than using an extinction curve, an attenuation curve should be used instead, see \citet{Fischera03,Fischera05} and Paper I.

The effect of an attenuation law will be to cause the strengths of lines in the red to be underestimated, while UV lines such as the \OIILs\ lines will be systematically overestimated. These effects will mean that the observations corrected using an extinction law will land systematically too far to the left and too low on this diagram, while on (almost) extinction-free diagnostics such as the V\&O plots, the data will plot onto the correct position. If this effect is the cause of the displacement between theory and observation seen in figure \ref{fig15}, this suggests that the  \OIIIL/\OIILs\ ratio may be systematically underestimated by $\sim 0.3$dex. This would imply that the  \NIIL/\OIILs\ ratio may be underestimated by a similar amount, since the extinction is determined from the \Ha/\Hb\ ratio which means that the observed \NIIL/\OIIIL\ ratio must necessarily be correct. The effect of a correction of this magnitude is shown on figure \ref{fig16} as an arrow marked `extinction'. With a correction of this magnitude, not only can the lower bound of the abundance in the NFGS survey be reconciled with that given by figure \ref{fig13} ($Z/Z_{\odot} \sim 0.2$), but also with the upper bound indicated by  figure \ref{fig15} ($Z/Z_{\odot} \sim 2$). The inferred range on $\log {\cal R}$ would then be 2 to -4.

A systematic overestimate of the  \OIILs\ lines would also lead to a systematic overestimate of \R23. This may well account for much of the discrepancy between theory and observation in figure \ref{fig6}.

\begin{figure}
 \includegraphics[width=\hsize]{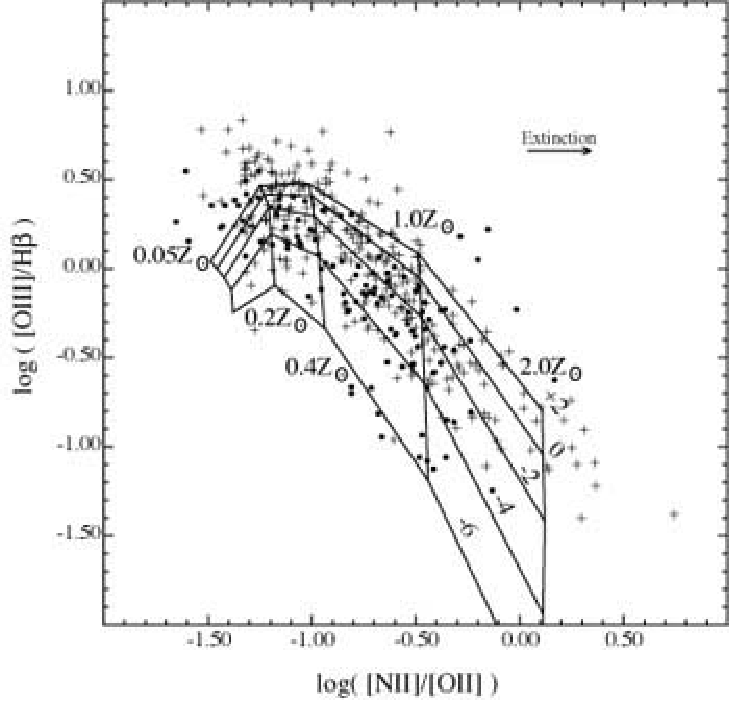}
\caption{\label{fig17}
 As figure \ref{fig13}, but for the \NIIL/\OIILs\ ratio against the \OIIIL/\Hb\ ratio used by \citet{Dopita00}. This should be compared with figure \ref{fig8}. This is a good diagnostic for the determination of both metallicity and $\log {\cal R}$. The observations and the theory would give consistent answers in all integral spectrum diagnostics if a systematic correction is made to the observations of the magnitude illustrated by the arrow marked `extinction'  in this and the previous figure (see text).}
\end{figure}

A similar effect is seen in the second  of the \citet{Dopita00} diagnostics is shown in figure \ref{fig17}. This plots the \NIIL/\OIILs\ ratio against the \OIIIL/\Hb\ ratio. Once again, a systematic extinction correction error of 0.3~dex would shift the observational points into a range which is consistent with the both the abundance and $R$ bounds inferred above.

\begin{figure}
 \includegraphics[width=\hsize]{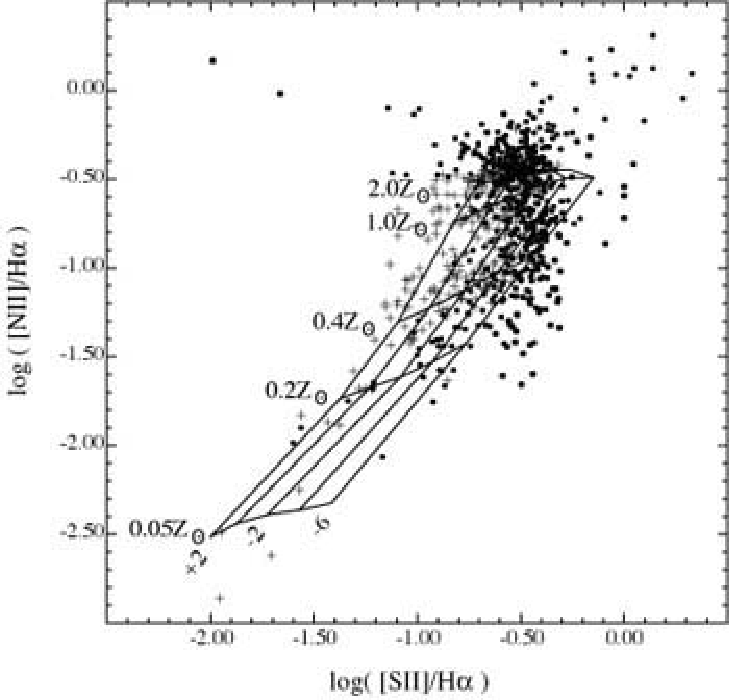}
\caption{\label{fig18}
 As figure \ref{fig13}, but for the \NIIL/\Ha\ ratio against the \SIILs/\Ha\ ratio. This should be compared with figure \ref{fig10}. This once again illustrates our problems in modelling the \SII\ line strengths.}
\end{figure}

Finally, we plot the \SIILs/\Ha\ ratio versus the  \NIIL/\Ha\ line ratio in figure \ref{fig18} (compare with figure \ref{fig10}). This shows, once again, the value of the \NIIL/\Ha\ line ratio as an abundance diagnostic in its own right. It also illustrates, once again, our difficulties with the \SIILs/\Ha\ ratio, with the observations, particularly the  \citet{Moustakas05} data set falling to the right of the theoretical curves in this ratio.

Note how the integrated spectra on figure \ref{fig18} are concentrated along the right-hand side of the region defined by the observations of single \HII\ regions. This reflects the fact that integrated galaxy spectra include the diffuse component of the ISM, even when this has a low surface brightness, because the total flux in the diffuse component may be an appreciable portion of the total emission. Indeed, a recent study by \cite{Oey06} suggests that the ``diffuse" fraction may exceed 50\% in nearby starburst galaxies.

\section{Discussion \& Conclusions}\label{discussion}

It should now be clear that these models represent a significant advance in our understanding of the strong line spectra of \HII\ regions. In particular, the models explicitly account for the dynamical evolution of the \HII\ region controlled through the mechanical energy input by both the stellar winds and supernova explosions. Two factors largely control the excitation of the young \HII\ regions, both of these dependent on the chemical abundance. These are:
\begin{itemize}
\item{The effective temperature of the exciting stars, which increases with decreasing metallicity, and}
\item{The control of the ionization parameter by the stellar winds in the expanding \HII\ regions. This was shown in Paper II to increase with decreasing metallicity}
\end{itemize}  
Since both of these effects operate to increase the excitation of low abundance \HII\ regions, we therefore find an abundance dependent excitation, which much improves the fit of the theory compared to models in which the ionization parameter is treated as a free variable. In our models the ionization parameter is replaced by the $R$ parameter, the ratio of the cluster mass in solar masses to the pressure in the ISM; $P/k$~(cm$^{-3}$K). This parameter is used because all models with a given $R$ have a unique relationship between their instantaneous ionization parameter and time. This allows us to construct evolutionary tracks for ensembles of \HII\ regions.

With increasing age, the ionization parameter falls almost monotonically, and the \HII\ region fades after reaching its maximum brightness in \Ha\ after 1-2~Myr. The presence of Wolf-Rayet stars after about 3.5~Myr causes the effective temperature of the cluster to briefly increase. This increases the excitation for a short time, causing the 3~Myr and the 4~Myr isochrones to cross over each other on diagnostic plots for the cases of high metallicity ($Z/Z_{\odot} > 0.5$, approximately).

In this paper, we have identified a number of useful diagnostics for deriving the abundance and either the age, in the case of isolated \HII\ regions, or the $R$ parameter, in the case of integrated spectra of galaxies. Our diagnostics for the integrated spectra are the first which explicitly take into account the fact that when we observe a whole galaxy, we are observing an ensemble average of \HII\ regions of all ages, sizes and central cluster masses. The simplification we have been able to make by using the $R$ parameter has rendered this problem tractable to a more simple analysis other than treating each \HII\ region separately, and adding the whole lot together at the end. 

In comparing the models to the observational data we have identified some potential problems with the way that observations of \HII\ regions have been made and analyzed. In particular:
\begin{itemize}
\item{Long-slit observations do not integrate over the whole of an \HII\ region, particularly in the case of observations of nearby, low metallicity dwarf irregular galaxies. By picking out the brightest regions of such \HII\ regions, we bias the data towards the high-excitation regions and the spectrum is not representative of the whole \HII\ region \citep{Luridana01,Stazinska03} and }
\item{The use of a reddening correction using an extinction curve, rather than an attenuation curve (which is appropriate for spatially extended objects) will cause the line intensities of UV lines to be systematically overestimated, biasing the strong line diagnostics which use the \OIILs\ \AA\ lines.}
\end{itemize}  

This second effect would reveal itself in the spectro-photometry as a failure to reproduce the correct intensities of the higher members of the Balmer Series, such as H$\delta$ or H$\epsilon$, and so should be quite easy to check for in observational data sets. Such an effect would also influence the measurement of the strength of the \OIII\ $\lambda 4363$\AA\ line, again causing it to be overestimated. This could be part of the reason for the long-standing discrepancy between the abundances derived from the strong lines and those obtained from measurements of the electron temperature using ratios such as \OIII\ $\lambda 4363$\AA/\OIIIL\ , see \citet{Bresolin04} for a recent discussion. We expect that the major part of the discrepancy will turn out to be caused by temperature fluctuations. However, this could be easily corrected for by assuming the Case B recombination line ratio between \Hb\ and H$\gamma$, and measuring the forbidden line strengths relative to the nearest Balmer lines in the spectrum.

There is clearly a need for integral field spectrophotometry on extragalactic \HII\ regions. This would allow efficient construction of integrated spectra and assist in the removal of stellar flux from the spectrum. Such observations would not only help to remove the remaining discrepancies between theory and observation but also give insight into the use of theoretical integral spectra in the analysis of the chemical evolution history of the early Universe.

Our models have their deficiencies. In general, all the low ionization species are predicted weaker than observed. These include the \OIL, the \SIILs, and the \OIILs, although these lines are probably affected by the reddening corrections made to the observations.

What could enhance the \SIILs/\Ha\ ratio or the \OIL/\Ha\ ratio? First, both the the \ion{O}{1} and the  \ion{S}{2} ion are very much more sensitive than the  \ion{N}{2} ion to the diffuse radiation field in \HII\ regions. They are therefore much enhanced in the vicinity of ionization fronts, and in shadow regions behind elephant trunk features, which are obscured from direct stellar radiation and illuminated only by the diffuse nebular radiation field. Second, both are much enhanced in shock regions. Indeed high  \SII/H$\alpha$ ratios are used as a means of identifying supernova remnants either embedded in \HII\ regions or in the ISM at large, and \OI/H$\alpha$ ratios reach 0.5 or higher in shock-excited regions. Neither of these effects are taken into account in the models.

In addition, the \ion{O}{1} ion is locked by charge-exchange reactions to the concentration of \HI. Furthermore, the collision strength of \OIL\ increases with electron temperature. To produce a high  \OIL/\Ha\ ratio in a purely photoionized region requires the presence of a hard radiation field. This could originate either as a thermal soft X-ray continuum from the shocked bubble of stellar wind material, or it could simply be an indication of a hotter sub-component in the global stellar radiation field.

If the radiation field produced by \emph{Starburst 99 v5.} is too soft, this could be rectified in at least three ways, First, by the use of a flatter IMF in the upper mass ranges ($M/M_{odot} > 10)$. This could be possible because observationally we have very few constraints other than by use of the excitation of \HII\ regions. Second, by adding another stellar component such as mass-exchange massive binaries to the stellar synthesis. 

A third - and quite likely possibility is that the EUV spectrum of the massive stars used in \emph{Starburst 99 v5.} may be in error. With previous versions of \emph{Starburst 99} based on the \emph{CoStar} models \citep{Schaerer97}, the EUV spectrum was, rather, too hard. However, \citet{Morisset04} has demonstrated that the theoretical number of ionizing photons emitted by the central star and the shape of the ionizing spectrum is highly dependent upon the atmospheric model used. At one extreme are the \emph{CoStar} models, at the the other extreme are found the `classical' \emph{Kurucz} \citep{Kurucz91, Kurucz94} plane-parallel LTE line blanketed models used in \emph{Starburts99 v5.}. The Kurucz models display the softest EUV spectra and the lowest photoionising flux, and are restricted to $\log g = 3.0$ for the models with higher effective temperature, $T_{\rm eff}$.  In between these extremes are found three sets of models with fairly similar photoinising flux predictions \citep{Morisset04}, the \emph{TLUSTY} models of \cite{Lanz03a, Lanz03b}, the \emph{WM-Basic} models of \cite{Pauldrach01} and the \emph{CMFGEN} models \cite{Hillier98}. Of these, the \emph{TLUSTY} models are plane-parallel hydrostatic, while the other two are spherical dynamic atmospheres. All of this offers plenty of potential for error in the estimation of the the hardness of the EUV spectrum for a given cluster mass and metallicity. 

Given that the \OIIIL/\Hb\ ratio of the models also appears to be on the low side, we believe that an increase in the stellar effective temperature, either by a flattening of the IMF or by some other means offers the best chance to more properly match the theory with the observations. The models are too weak in their forbidden line strengths overall. The \citet{Stoy33} method \citep{Pottasch83} of deriving the effective temperature of the exciting stars relies upon measuring the total flux cooling lines (which is a measure of the mean energy per photoionization) to a recombination line (which counts the number of photoionizations). Thus, if we want to increase the absolute strengths of the forbidden lines, we have to increase the stellar effective temperature. This increase would also strongly increase the  \OIL/\Ha\ ratio by both increasing the extent of the partially-ionized zonr and by increasing the electron temperature within it.

A niggling uncertainty with the modeling is the mis-match between the \emph{Starburst 99 v5} abundance set and the abundance set given in table \ref{table1}. The \emph{Starburst 99 v5} has to rely on the older stellar evolution models which use the old value of the solar abundances. In an attempt to quantify this effect, we ran a test model with a $0.4Z_{\odot}$ spectral synthesis cluster model from \emph{Starburst 99 v5} embedded in a nebula with $1.0Z_{\odot}$ (using the abundances given in table \ref{table1}). This made surprisingly little difference, changing critical line ratios by 0.1~dex or less, except for the \OIL/\Ha\ ratio, which increased by 34\%. We therefore conclude that the results presented in this paper are reasonably secure against any future adjustment of stellar atmospheric abundances.

In conclusion, the major advances in modelling made in this paper are:
\begin{itemize}
\item{For individual HII regions, it becomes possible to estimate the ages  of the exciting stars from the positions of the observations on HII region isochrones. Previously, this was not possible, since the
ionization parameter and the effective temperature of the cluster  (determined by the ageing of its stars) produce similar effects on the  emitted spectrum. The self-consistent treatment  we have developed which  includes the dynamical evolution of the HII region consistent with the properties of the stellar wind generated by the central cluster, and which recognises that the stellar wind determines the pressure 
in the HII region allows us to eliminate the use of arbitrary geometries.}
\item{ In addition, this work gives a methodology to construct much more ``realistic" integrated spectra of HII galaxies, something that has not been possible before in the approximations which assume that a whole galaxy can be represented by a single spherical \HII\ region with a chosen abundance and ionization parameter.}
\end{itemize}

 \begin{acknowledgements}
Dopita acknowledges the support of both the Australian National University and  the Australian Research Council (ARC) through his ARC Australian Federation Fellowship. Dopita, Sutherland \& Fishera acknowledge financial support of ARC Discovery project grant DP0208445. Kewley acknowledges a Hubble Fellowship. The work by van Breugel was performed under the auspices of the U.S. Department of Energy and Lawrence Livermore National Laboratory under contract No. W-7405-Eng-48.

\end{acknowledgements}

\appendix
Here we give a summary of the strong line flux ratios with respect to \Hb\ as computed in the photoionization models. The line identifications, the ionic species and the wavelengths of each line listed in the electronic version of this paper are given in Table 2.

In the electronic version of the paper, separate tables are given for each of the abundances computed; 0.05, 0.2, 0.4, 1.0 and 2.0 $Z_{\odot}$, and in each of these tables, line fluxes with respect to \Hb\ $= 1.0$ are given for each of the 51  lines listed in Table 2. These spectral line intensities are given for each value of $\log {\cal R}$ computed; +2, 0, -2, -4 and -6. The quantity $R$ is that defined in the text of the paper:

\begin{equation}
 {\cal R} = \left[\frac{M_{\rm cl}}{M_{\odot}}\right] \left[\frac{P/k}{\rm 10^4 cm^{-3}K}\right]
\end{equation}

The electronic tables (Tables 3 through 7) will be available when the paper is published.

\begin{deluxetable}{cccccc}
\tabletypesize{\small}
\tablewidth{0pt}
\tablecaption{Line ID and Wavelength list}
\tablehead{\colhead{Line ID \#} & \colhead{Ion} & \colhead{$\lambda$ (\AA)}& \colhead{Line ID \#} & \colhead{Ion} & \colhead{$\lambda$ (\AA)}  \\ }
\startdata
01 & \ion{H}{1} & 1215.7  & 27 & [\ion{S}{2}] & 6716.3\\
02 & \ion{C}{3}] & 1909.1 & 28 & [\ion{S}{2}] & 6730.7\\
03 & \ion{C}{3}] & 1911.2 & 29	& [\ion{Ar}{3}] &7135.7\\
04 & [\ion{C}{2}] & 2325.2 & 30	& [\ion{Ar}{3}] &7751.0\\
05 & \ion{Mg}{2} & 2797.9 & 31 & [\ion{S}{3}] & 9069.3\\
06 & [\ion{O}{2}] & 3726.0 & 32 & \ion{H}{1} & 10049.4\\
07 & [\ion{O}{2}] & 3728.7 & 33 & \ion{H}{1} & 10049.4\\
08 & \ion{H}{1} & 3797.9 & 34 & \ion{He}{1} & 10830.0\\
09 & \ion{H}{1} & 3835.4 & 35 & \ion{He}{1} & 10833.0\\
10 & [\ion{Ne}{3}] & 3868.7 & 36 & \ion{H}{1} & 10938.1\\
11 & \ion{He}{1} & 3888.6 & 37 & \ion{H}{1} & 12818.1\\
12 & \ion{H}{1} & 3889.1 & 38 & \ion{H}{1} & 18751.0\\
13 & [\ion{Ne}{3}] & 3967.4 & 39 & \ion{H}{1} & 26252.0\\
14 & \ion{H}{1} & 3970.1 & 40 & \ion{H}{1} & 40512.0\\
15 & \ion{H}{1} & 4104.7 & 41 & [\ion{Ar}{2}] & 69832.8\\
16 & \ion{H}{1} & 4340.5 & 42 & [\ion{Ar}{3}] & 89892.5\\
17 & \ion{He}{1} & 4471.5 & 43 & [\ion{S}{4}] & 105221\\
18 & \ion{H}{1} & 4861.3 & 44 &[\ion{Ne}{2}] & 128115\\
19 & [\ion{O}{3}] & 4958.8 & 45 & [\ion{Ne}{3}] & 155513 \\
20& [\ion{O}{3}] & 5006.8 & 46 & [\ion{S}{3}] & 186821\\
21& \ion{He}{1} & 5875.6 & 47 & [\ion{S}{3}] & 336366\\
22 & [\ion{O}{1}] & 6300.2 & 48 & [\ion{Si}{2}] & 347941\\
23 & [\ion{N}{2}] & 6548.0 & 49 & [\ion{O}{3}] & 517972\\
24 & \ion{H}{1} & 6562.8 & 50 & [\ion{N}{3}] &573845\\
25 & [\ion{N}{2}] & 6583.3 & 51 & [\ion{O}{3}] & 883017\\
26 & \ion{He}{1} & 6678.2 & .. & .. & .. \\
\enddata
\end{deluxetable}

\setlength{\voffset}{0mm}

\end{document}